\documentclass[a4paper,12pt]{article}
\usepackage{amsmath}
\usepackage{amssymb,mathrsfs}
\usepackage{mathtext}
\usepackage[english]{babel}
\usepackage[left=1cm,right=1cm,top=1cm,bottom=3cm]{geometry}
\usepackage{graphicx}
\usepackage{subfigure}
\usepackage{mathtools}
\usepackage{geometry}
\usepackage{authblk}

\newcommand{\prt}{\partial}

\graphicspath{{figures/}}

\title{Expansion Dynamics of a Two-Component Quasi-One-Dimensional
Bose-Einstein Condensate: Phase Diagram, Self-Similar Solutions,
and Dispersive Shock Waves}

\author[1,2]{S.~K. Ivanov}
\author[1]{A.~M. Kamchatnov}
\affil[1]{Institute of Spectroscopy, Russian Academy of Sciences,
Russia, Moscow, Troitsk, 108840,}
\affil[2]{Moscow Institute of Physics and Technology, Russia, Moscow, 117303}
\date{September 20, 2016}

\date{}

\begin{document}

\maketitle

\abstract{We investigate the expansion dynamics of a Bose-Einstein condensate that consists of two components
and is initially confined in a quasi-one-dimensional trap. We classify the possible initial states of the
two-component condensate by taking into account the non-uniformity of the distributions of its components
and construct the corresponding phase diagram in the plane of nonlinear interaction constants. The differential
equations that describe the condensate evolution are derived by assuming that the condensate density
and velocity depend on the spatial coordinate quadratically and linearly, respectively, what reproduces the
initial equilibrium distribution of the condensate in the trap in the Thomas-Fermi approximation. We
obtained self-similar solutions of these differential equations for several important special cases and wrote out
asymptotic formulas describing the condensate motion on long time scales, when the condensate density
becomes so low that the interaction between atoms can be neglected. The problem on the dynamics of
immiscible components with the formation of dispersive shock waves was also considered. We compare the numerical
solutions of the Gross-Pitaevskii equations with their approximate analytical solutions and study numerically
the situations when the analytical method admits no exact solutions.}

\section{Introduction}

The dynamics of a Bose-Einstein condensate is
the subject of active current research. A multitude of
experimental and theoretical works aimed at studying
the solitons, vortices, dispersive shock waves, and
other structures that determine the characteristic features
of the behavior of a condensate in various experimental
conditions have been performed by now (see,
e.g.,~\cite{PitaevskiiStringari-03}). One of the main problems referring to this
direction of research is to study the expansion dynamics
of a condensate after the trap confining this condensate
has been switched off, because in many experiments
the results are recorded after the condensate
expansion to a state when the cloud sizes are large
enough for the measurements to be made (see, e.g.,
the experiments in~\cite{AndrewsMDDK-96,ErnstSSMKR-98}). This problem was first
investigated theoretically in \cite{KaganSS-96} in the hydrodynamic
approximation, where the equations admit a simple
self-similar solution. This approach was then developed
in \cite{CastinDum-96,DalfovoMSP-97,KaganSS-97,Kamchatnov-04} for a condensate consisting of one component,
and good agreement between theory and
experiment was found. However, the situation changes
significantly for the case of a condensate consisting of
several components, where, for example, atoms of two different species (see~\cite{FerrariIJMRS-02,ModugnoMRR-02}), two different isotopes of
one species of atoms (see~\cite{PappPW-08}), or one species of atoms
in two different quantum states, such that the difference
of the energy levels of these states is much smaller
than the condensate temperature (see~\cite{DelannoyMBJBA-01,MyattBGCW-97,HallMEWC-98}), are
condensed. In particular, in two-component condensates
the cases of relatively strong mutual repulsion
between the components, where they are immiscible,
and relatively weak mutual repulsion, where they are
miscible, i.e., occupy the same volume, should be distinguished.
This difference between the condensate
phase states affects significantly the dynamics of the
condensate, including the dynamics of its expansion.
So far this dynamics has not been studied comprehensively
enough. Some partial results illustrating the difference
between the expansions of one-component
and two-component condensates were obtained in \cite{Quach-14}. However, in this paper the author used predominantly
numerical methods. In our paper we will show
that there are interesting situations where a comprehensive
analytical study can be carried out within the
hydrodynamic approximation used previously in the
one-component case in \cite{KaganSS-96,CastinDum-96,DalfovoMSP-97,KaganSS-97,Kamchatnov-04}. Comparison with
numerical calculations shows that although the dispersion effects can play some role at intermediate
expansion stages, nevertheless, these effects become
extremely small and they may be neglected at the
asymptotic stage, which is most interesting for an
experiment. The assumption that the evolution of the
component density and velocity profiles is self-similar,
which generalizes the approach from \cite{KaganSS-96}, plays a
significant role in such cases favorable for the analytical
theory. However, we will show that this assumption
does not always adequately describe the dynamics if
the initial state of the condensates before their release
from the trap is near the boundary of the phase transition
between component miscibility and immiscibility,
and the problem requires a numerical solution in
this case. Nevertheless, even in the case of immiscible
components one can distinguish a characteristic case
where one of the components may be considered as a
piston moving the other component. For such an idealized
situation the condensate expansion is accompanied
by the formation of a dispersive shock wave in one
component and a rarefaction wave in the other component.
The theory developed for this case agrees well
with the numerical results. The results of this paper
allow the characteristic features of the phenomenon
depending on the condensate parameters to be predicted.

\section{The hydrodynamical form of the Gross-Pitaevskii equations}

The dynamics of a Bose-Einstein condensate
under the action of a potential $U$ is described with a
high accuracy by the Gross-Pitaevskii equations. In
the two-component case that we will be concerned
with here, these can be written as
\begin{equation} \label{GPETC}
    \begin{split}
    i\hbar\frac{\partial\psi_i}{\partial t}=-\frac{\hbar^2}{2m_i}\Delta\psi_i+g_{ii}|\psi_i|^2\psi_i+g_{ij}|\psi_{j}|^2\psi_i+U_i\psi_i,
    \end{split}
\end{equation}
where $i,j=1,2~(i\neq j)$ is the number of the corresponding
condensate component, $(\psi_1,\psi_2)$ are the wave
functions of the components,
$g_{ii}$ are the interaction
constants between atoms of component $i$, and $g_{ij}$ are
the interaction constants between atoms of different
species. Usually, $g_{12}=g_{21}$, which we will assume in the
subsequent discussion. The interaction constants can
be expressed via the scattering lengths $a_{ij}$ of atoms by
one another as
    \begin{eqnarray} \label{g}
    g_{ii}=\frac{4\pi\hbar^2a_{ii}}{m_i},\qquad g_{ij}=\frac{2\pi\hbar^2a_{ij}}{m_{ij}},
    \end{eqnarray}
where $m_{ij}^{-1}=m_i^{-1}+m_j^{-1}$ is the reduced mass of the
atoms being scattered by one another. Each of the
wave functions is normalized to the number of particles
of a given species in the condensate:
    \begin{eqnarray} \label{Norm}
    \int|\psi_i|^2dV=N_i,
    \end{eqnarray}
so that $|\psi_i|^2=\rho_i$ is the number density of atoms in the $i$th component. The gradient of the phase $\varphi_i$ of the wave
function $\psi_i=\sqrt{\rho_i}\exp(i\varphi_i)$ is related to the flow velocity $\textbf{u}_i$ of $i$th component by
the relation (see~\cite{PitaevskiiStringari-03})
    \begin{equation} \label{velocity}
    \textbf{u}_i=\frac{\hbar}{m_i}\nabla\varphi_i.
    \end{equation}

The condensate components are miscible, i.e.
their uniform distribution over space (in the absence
of an external potential) is stable, if the interaction
constants satisfy the condition (see~\cite{AoChui-98})
    \begin{equation} \label{gg}
    g_{12}^2<g_{11}g_{22}.
    \end{equation}
If, alternatively, the sign of this inequality is opposite,
then the condensate is unstable with respect to the
separation into regions containing the components of
only one of the condensate species. However, this
condition is valid only for a uniform distribution of the
condensate. In contrast, for the case of a condensate
confined in a trap, the miscibility condition requires a
modification, which we will dwell on in more detail in
the next section of our paper.

If the phase $\varphi_i$ is a single-valued function of coordinates,
which physically means the absence of vortices
in the condensate, then the wave functions of the
two-component condensate can be represented as
    \begin{eqnarray} \label{WFtc}
    \psi_i=\sqrt{\rho_i(\textbf{r},t)} \exp\left(i\frac{m_i}{\hbar}\int^{\textbf{r}}\textbf{u}_i(\textbf{r}',t)d\textbf{r}'-i\frac{\mu_i}{\hbar}t\right),
    \end{eqnarray}
where $\mu_i$ is the chemical potential of the $i$th component
(see~\cite{PitaevskiiStringari-03}).
Substituting (\ref{WFtc}) into (\ref{GPETC}), separating the
real and imaginary parts, and differentiating one of the
equations with respect to $r$ bring the Gross-Pitaevskii
equations to the so-called hydrodynamic form:
    \begin{equation} \label{CE1}
    \frac{\partial\rho_i}{\partial t}+\nabla(\rho_i\mathbf{u}_i)=0,
    \end{equation}
    \begin{equation} \label{EEwqs}
    \begin{split}
    \frac{\partial \mathbf{u}_i}{\partial t}&+(\mathbf{u}_i\nabla) \mathbf{u}_i+ \frac{g_{ii}}{m_i}\nabla\rho_i
    +\frac{g_{ij}}{m_i}\nabla\rho_j+\frac{\nabla U_i}{m_i}+ \frac{\hbar^2}{4m_i^2}\nabla\left(\frac{(\nabla\rho_i)^2}{2\rho_i^2}
    -\frac{\Delta\rho_i}{\rho_i}\right)=0.
    \end{split}
    \end{equation}
Equations (\ref{CE1}) are responsible for the conservation of
the number of particles in the corresponding condensate
component. If there were no last term proportional
to $\hbar^2$ in Eqs.~(\ref{EEwqs}), then these equations would
correspond to ordinary Eulerian hydrodynamics with
a pressure gradient $\nabla p_i=(g_{ii}\nabla\rho_i+g_{ij}\nabla\rho_j)/m_i$.
However, the
last term of Eqs.~(\ref{EEwqs}) attributable to the dispersion of
quantum particles introduces new properties if the
condensate characteristics change rapidly enough. Let
us make an estimate for the distance $\xi$ at which the
pressure and dispersion contribute identically. We
assume that the masses of atoms and the number densities
of particles in the components are of the same order of magnitude ($m_i\sim m_j$, $\rho_i\sim\rho_j$) for both components,
so that by $m$, $\rho$ and $g$ we can understand the
corresponding parameter of any component. We will
then estimate the pressure as $p\sim g\rho^2/2m$ and obtain $g\rho\sim\hbar^2/(m\xi^2)$,
for $\xi$, whence $\xi\sim\hbar/\sqrt{g\rho m}$. Thus, the
condensate has an intrinsic characteristic size that is
called the correlation length and can be defined as
    \begin{eqnarray} \label{Correlation}
    \xi\sim\frac{\hbar}{\sqrt{g\rho m}}.
    \end{eqnarray}
If the characteristics change weakly at distances $\sim\xi$, then the last term in Eqs.~(\ref{EEwqs}) can be neglected, and
the system will then take the form
    \begin{eqnarray} \label{CE}
    \frac{\partial\rho_i}{\partial t}+\nabla(\rho_i\mathbf{u}_i)=0,
    \end{eqnarray}
    \begin{eqnarray} \label{EE}
    \frac{\partial \mathbf{u}_i}{\partial t}+(\mathbf{u}_i\nabla) \mathbf{u}_i
    +\frac{g_{ii}}{m_i}\nabla\rho_i +\frac{g_{ij}}{m_i}\nabla\rho_j+\frac{\nabla U_i}{m_i}=0.
    \end{eqnarray}
Equations (\ref{EE}) correspond to Eulerian hydrodynamics,
and this form of hydrodynamic equations
describes fairly smooth solutions, in particular, the
component density distributions in the trap before it is
switched off. If, however, solitions with sizes $\sim\xi$, are
generated or dispersive shock waves are formed in the
course of evolution, then the dispersion effects should
also be taken into account. Such a case will be considered
in Section \ref{SectionDSW}.

To describe the characteristic features of the phenomenon,
we will dwell on the example of traps in
which the motion of particles in two directions is "frozen"
and is reduced to zero-point oscillations. In an
experiment such a quasi-one-dimensional condensate
acquires a highly elongated cigar shape. The potential
of such a trap for the $i$th component can be written as
    \begin{eqnarray} \label{U}
    U_i=\frac{1}{2}m_i[\omega_x^2x^2+\omega_\bot^2(y^2+z^2)],
    \end{eqnarray}
where $\omega_y=\omega_z\equiv\omega_\bot\gg\omega_x$. Owing to the latter inequality,
the motion of the condensate in the transverse
direction is frozen, i.e., the transverse wave function is
reduced to the ground state in a transverse potential
with frequency $\omega_\bot$. The Gross-Pitaevskii equation
can then be averaged over the transverse direction, and
the dynamics of the condensate is reduced to its
motion in the longitudinal $x$ direction (for details, see,
e.g.,~\cite{Kamchatnov-04}). Introducing the effective nonlinear constants
of longitudinal condensate dynamics (but
retaining, for simplicity, the previous notation for
them), we arrive at the equation
    \begin{equation} \label{GPEtcod}
    \begin{split}
    i\hbar\frac{\partial\psi_i}{\partial t}=-\frac{\hbar^2}{2m_i}
    \frac{\partial^2\psi_i}{\partial x^2}+g_{ii}|\psi_i|^2\psi_i+g_{ij}|\psi_{j}|^2\psi_i+\frac{1}{2}m_i\omega_i^2x^2\psi_i,
    \end{split}
    \end{equation}
while Eqs. (\ref{CE}) and (\ref{EE}) will transform to
    \begin{eqnarray} \label{CE1D}
    \frac{\partial\rho_i}{\partial t}+\frac{\partial}{\partial x}(\rho_iu_i)=0,
    \end{eqnarray}
    \begin{eqnarray} \label{EE1D}
    \frac{\partial u_i}{\partial t}+u_i\frac{\partial u_i}{\partial x}+ \frac{g_{ii}}{m_i}\frac{\partial\rho_i }{\partial x} +\frac{g_{ij}}{m_i}\frac{\partial\rho_j}{\partial x}+\omega_i^2x=0.
    \end{eqnarray}
Having established the basic equations for the dynamics
of a binary condensate, let us first turn to the problem
of classifying the possible initial density distributions
of the components before their release from the
trap.

\section{The phase diagram for a binary condensate confined in a quasi-one-dimensional trap} \label{SectionPD}

\begin{figure} \centering
    \includegraphics[width=0.5\linewidth]{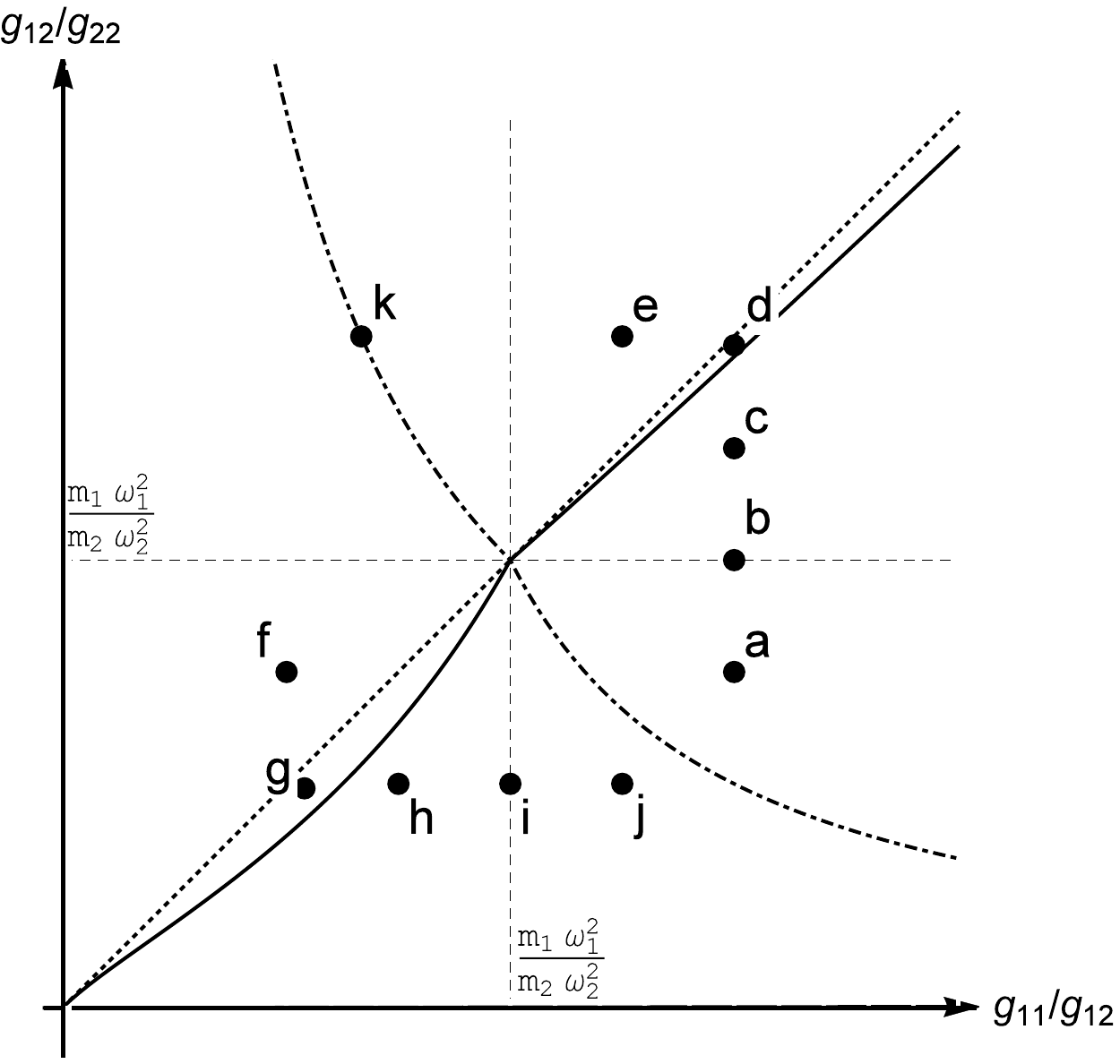}
    \caption{\textit{Phase diagram of the stationary particle number density distribution in the Thomas-Fermi approximation. The points on the diagram are marked by the letters corresponding to the graphs in Fig.~\ref{TomasFermi}. The diagram was constructed for condensates with identical masses of atoms and trap frequencies. The number of particles in the second component is twice that in the first one. The same parameters were also taken for the graphs in Fig.~\ref{TomasFermi}.}}
    \label{PhaseDiag}
    \end{figure}

\begin{figure}[t] \centering
        \subfigure[]{\includegraphics[width=0.2\linewidth]{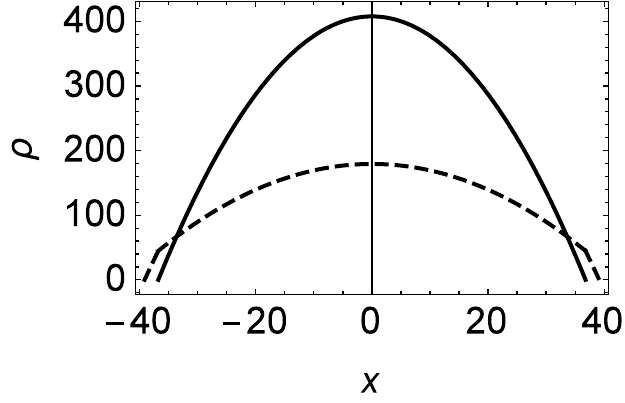} \label{TomasFermiA}}
        \subfigure[]{\includegraphics[width=0.2\linewidth]{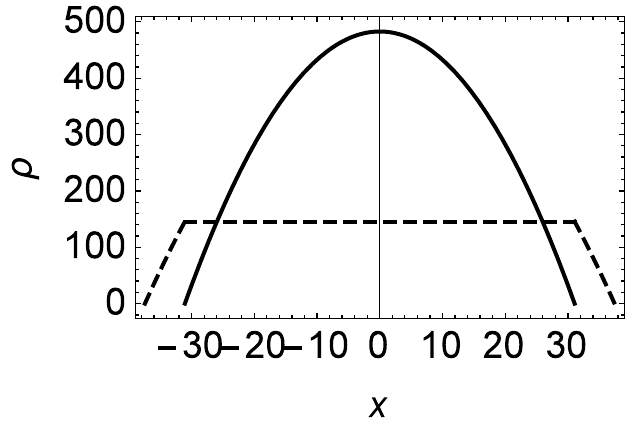} \label{TomasFermiB}}
        \subfigure[]{\includegraphics[width=0.2\linewidth]{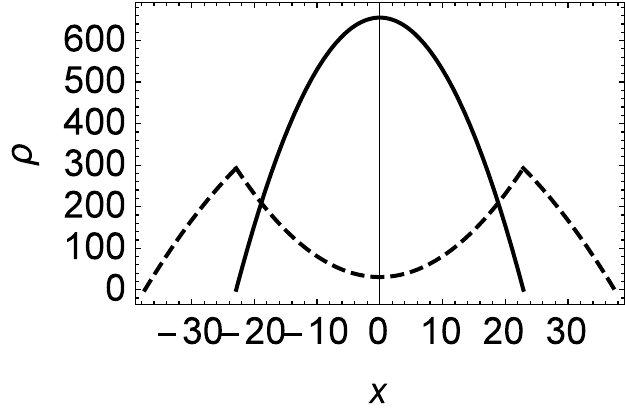} \label{TomasFermiC}}
        \subfigure[]{\includegraphics[width=0.2\linewidth]{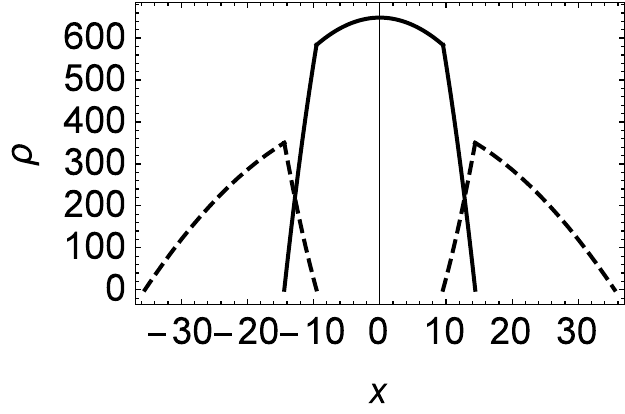} \label{TomasFermiD}}
        \subfigure[]{\includegraphics[width=0.2\linewidth]{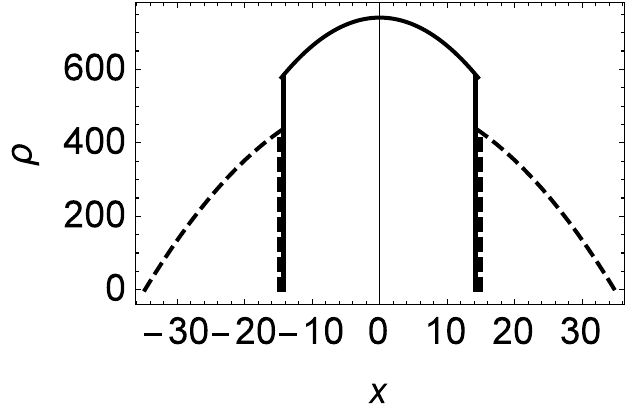} \label{TomasFermiE}}
        \subfigure[]{\includegraphics[width=0.2\linewidth]{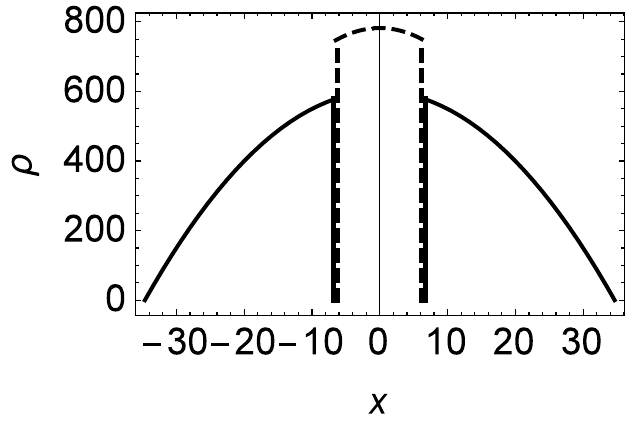} \label{TomasFermiF}}
        \subfigure[]{\includegraphics[width=0.2\linewidth]{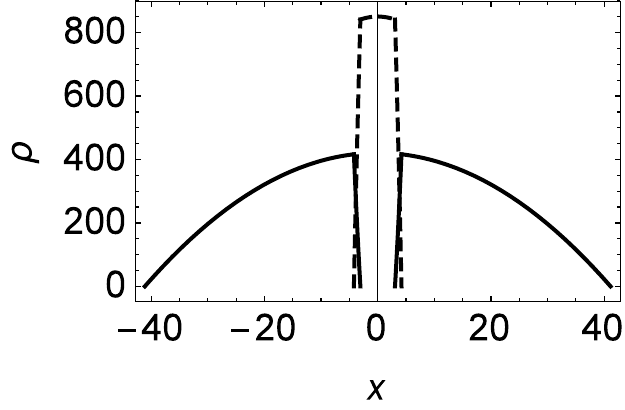} \label{TomasFermiG}}
        \subfigure[]{\includegraphics[width=0.2\linewidth]{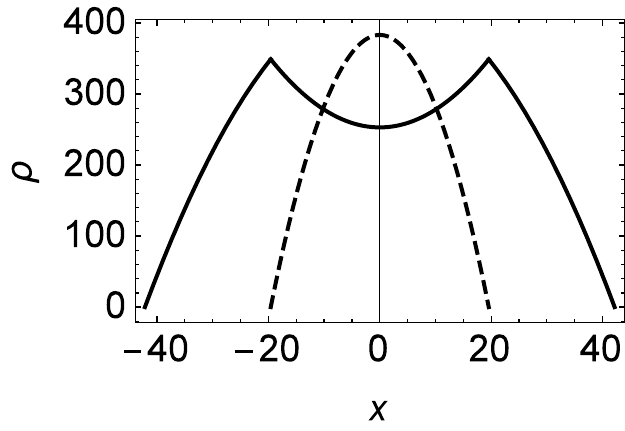} \label{TomasFermiH}}
        \subfigure[]{\includegraphics[width=0.2\linewidth]{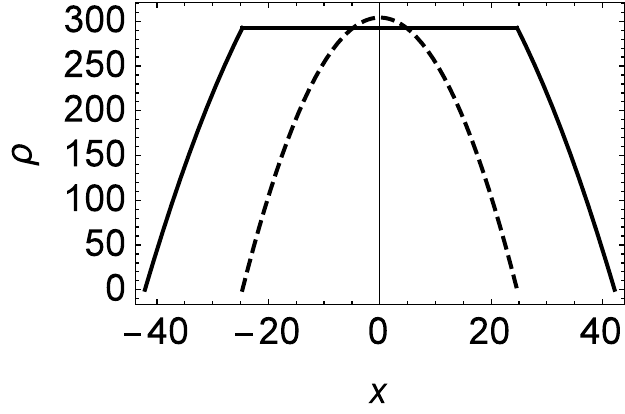} \label{TomasFermiI}}
        \subfigure[]{\includegraphics[width=0.2\linewidth]{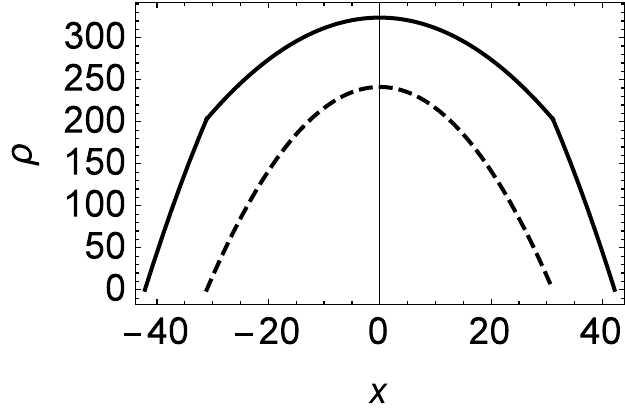} \label{TomasFermiJ}}
        \subfigure[]{\includegraphics[width=0.2\linewidth]{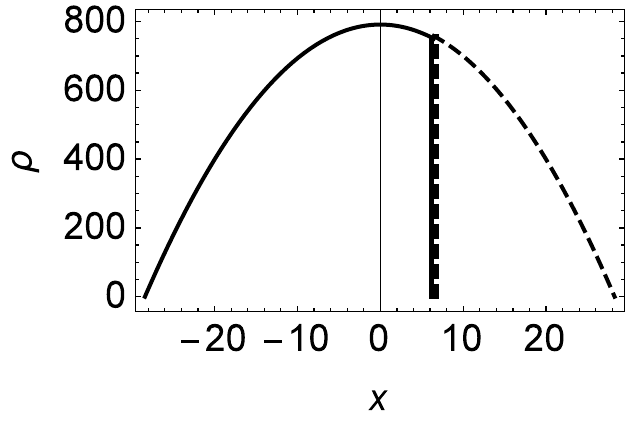} \label{TomasFermiK}}
    \caption{\textit{Typical particle number density profiles corresponding to different regions of the phase diagram
    Fig.~\ref{PhaseDiag}. The dashed and solid lines represent the first and second components, respectively. The miscibility phase is reflected by profiles \subref{TomasFermiA}-\subref{TomasFermiC} and \subref{TomasFermiH}-\subref{TomasFermiJ}, the symmetric immiscibility phase is reflected by profiles \subref{TomasFermiD}-\subref{TomasFermiG}, and the asymmetric immiscibility phase is reflected by profile \subref{TomasFermiK}. The point in Fig.~\ref{PhaseDiag} marked by the same letter as the graph corresponds to each graph.}}
    \label{TomasFermi}
    \end{figure}

Numerical calculations (see, e.g.,~\cite{LeeJKWAP-16,TrippenbachGRMB-00}) and experimental works (see, e.g.,~\cite{TojoTMHSH-10})
show that various
particle number density profiles can be realized,
depending on the relation between the interaction
constants, particle masses, numbers of particles in
each component, and trap frequencies. Obviously, the
condensate loaded into a trap will be distributed over
space so as to minimize the total energy
    \begin{equation} \label{Hamil}
    \begin{split}
    H=&\int \Big[\frac{\hbar^2}{2m_1}|\nabla\psi_1|^2+\frac{\hbar^2}{2m_2}|\nabla\psi_2|^2 +\frac{1}{2}(g_{11}|\psi_1|^4+2g_{12}|\psi_1|^2|\psi_2|^2+g_{22}|\psi_2|^4)+\\
    &+\frac12m_1\omega_1^2x^2|\psi_1|^2+\frac12m_2\omega_2^2x^2|\psi_2|^2\Big]dx.
    \end{split}
    \end{equation}
These distributions can have different forms, depending
on the nonlinear constants and trap frequencies,
and in this section we will classify the possible forms in
the Thomas-Fermi approximation, where the dispersion
properties of the condensate may be neglected:
    \begin{equation}
    \begin{split}
    H=\frac{1}{2}\int(g_{11}\rho_1^2+2g_{12}\rho_1\rho_2+g_{22}\rho_2^2+m_1\omega_1^2x^2\rho_1+m_2\omega_2^2x^2\rho_2)dx.
    \end{split}
    \end{equation}
This approximation will allow us to establish the main
types of possible distributions on a qualitative level.
For the Thomas-Fermi approximation to be applicable,
the size of each condensate cloud must be much
greater than the correlation length $\xi$ what we will
assume below.
First of all, note that in the distribution there can
be regions of space where both components are present
(``overlap'' regions) and regions where only one of
the components is present (``singlet'' region). Therefore,
let us write out the stationary solution of Eqs. (\ref{CE1D}) and (\ref{EE1D}) that corresponds to the Thomas-Fermi
approximation for these two possible cases:
    \begin{eqnarray}
    \label{IDfo}
    &&\rho_i^o(x)=\frac{2(g_{jj}\mu_i-g_{ij}\mu_j)-(g_{jj}m_i\omega_i^2-g_{ij}m_j\omega_j^2)x^2}{2(g_{ii}g_{jj}-g_{ij}^2)}, \\
    \label{IDfs}
    &&\rho_i^s(x)=\frac{2\mu_i-m_i\omega_i^2x^2}{2g_{ii}},
    \end{eqnarray}
where $\rho_i(x)=|\psi_i(x,0)|^2$. Here, the index ``$o$'' (overlap)
denotes the particle number density in the overlap
region and the index ``$s$'' (singular) denotes the densities
in the singlet regions where only one of the components
is located. The chemical potentials $\mu_i$ are
functions of the number of particles, particle masses,
interaction constants, and trap frequencies. These
functions are defined by the equations
    \begin{equation} \label{NormRo}
    \int\rho_1dx=N_1, \qquad
    \int\rho_2dx=N_2,
    \end{equation}
where $N_1$ and $N_2$ are the numbers of particles in the
first and second components, respectively, and the
integration is over the region where the corresponding
condensate component is located. It is easy to find the
sizes of each of the components of the condensate
confined in the trap from Eqs.~(\ref{IDfo}) and (\ref{IDfs}).
Following \cite{LeeJKWAP-16}, we will classify the possible configurations by
associating them with points on the plane with coordinate
axes $(g_{11}/g_{12},g_{12}/g_{22})$. These variables characterize
the relative value of the interaction constants. The
phase diagram arising in this way is shown in Fig.~\ref{PhaseDiag}, while the typical distributions corresponding to the
points on this plane are shown in Fig.~\ref{TomasFermi}. Let us introduce
the following terminology for the various phases
that can be identified in these figures. We will call the
configuration where both components overlap at the trap center the miscibility phase (Fig.~\ref{TomasFermi}\subref{TomasFermiA}-\subref{TomasFermiC},\subref{TomasFermiH}-\subref{TomasFermiJ}) and the configuration where the components are separated
and one of the components is surrounded by
the other the symmetric immiscibility phase (Fig.~\ref{TomasFermi}, \subref{TomasFermiD}-\subref{TomasFermiG}).
The lines in Fig.~\ref{PhaseDiag} separate the regions
by the following attributes. It follows from (\ref{gg}) that the
diagonal $g_{11}/g_{12}=g_{12}/g_{22}$ (dotted line)
separates the
miscibility/immiscibility regions of a homogeneous
condensate: the components in the homogeneous
condensate are immiscible above the diagonal and
miscible below it. Because of the influence of the trap,
this line now plays a slightly different role---it separates
the condensates with nonzero (Fig.~\ref{TomasFermiD} and \ref{TomasFermiG}) and zero
(Fig.~\ref{TomasFermiE} and \ref{TomasFermiF}) widths of the overlap region.
These two sets of figures differ by the numbers of the
components located at the trap center and outside the
central component (in the ``shell'').

On the solid lines the density of the external component
becomes zero at the trap center, i.e., according
to our definition, these lines separate the configurations
with miscibility and immiscibility. The equation
for the curves on which the first component forms an
external shell and its density becomes zero at the
trap center is analytically expressed by the formula
$g_{12}/g_{22}=\mu_1/\mu_2$ which can be derived from the condition $\rho_1^o(0)=0$,
and the other branch for which the first
component is internal and the second one is external
is defined in a similar way. The ratio of the chemical
potentials can be found from system (\ref{NormRo}), and we
obtain the following equation for the first component
to become zero:
    \begin{eqnarray}
    \frac{g_{12}}{g_{22}}=\frac{\frac{N_1}{N_2}\frac{g_{11}}{g_{12}}-\frac{1}{2}+
    \sqrt{(1+\frac{N_1}{N_2}\frac{m_1\omega_1^2}{m_2\omega_2^2})\frac{N_1}{N_2}
    \frac{g_{11}}{g_{12}}+\frac{1}{4}}}{(\frac{N_1}{N_2})^2\frac{m_1\omega_1^2}{m_2\omega_2^2}
    +2\frac{N_1}{N_2}-(\frac{N_1}{N_2})^2\frac{g_{11}}{g_{12}}}.
    \end{eqnarray}
The second component becomes zero at the trap center
when passing through the curve
    \begin{eqnarray}
    \frac{g_{12}}{g_{22}}=\frac{(\frac{N_1}{N_2}\frac{g_{11}}{g_{12}}+1)^2}{(\frac{N_1}{N_2})^2\frac{g_{11}}{g_{12}}
    +\frac{m_2\omega_2^2}{m_1\omega_1^2}+2\frac{N_1}{N_2}}.
    \end{eqnarray}
The solid lines in Fig.~\ref{PhaseDiag} indicate examples of the
curves, where the external component at the center of
symmetry becomes zero in the Thomas-Fermi
approximation, for the number of particles in the second
component that is twice that in the first one and
identical masses and trap frequencies. The passage
through these lines is illustrated by a qualitative difference
between the distributions in Figs.~\ref{TomasFermiG}
and \ref{TomasFermiH} (the
first component at the center) and Fig.\ref{TomasFermiD} and \ref{TomasFermiC}
(the second
component at the center).

The distributions in \ref{TomasFermiC} and \ref{TomasFermiH} for the external
components are concave at the trap center. However, as one recedes from the solid curves, the distributions
of the external components at the center become flatter
and, at some moment, they become horizontal.
The equation $\partial\rho_1^o(x)/\partial x=0$ gives the condition for the
first component being constant, while the equation $\partial\rho_2^o(x)/\partial x=0$ gives the condition for the second component
being constant. Accordingly, we find the relations
between the constants of our problem:
    \begin{eqnarray}\label{polochka}
    \frac{g_{11}}{g_{12}}=\frac{m_1\omega_1^2}{m_2\omega_2^2}\quad\text{or}\quad
    \frac{g_{12}}{g_{22}}=\frac{m_1\omega_1^2}{m_2\omega_2^2}.
    \end{eqnarray}
These straight lines are indicated in Fig.~\ref{PhaseDiag} by the
dashed lines parallel to the coordinate axes. In particular,
below the line $g_{12}/g_{22}=m_1\omega_1^2/m_2\omega_2^2$ the first component
has an upward-convex distribution (the dashed
line in
Fig.~\ref{TomasFermiA}) in the overlap region, the distribution
becomes flat on this line (Fig.~\ref{TomasFermiB}),
and slightly above
this line it becomes downward-concave (Fig.~\ref{TomasFermiC}).

\begin{figure}[t] \centering
        \subfigure[]{\includegraphics[width=0.45\linewidth]{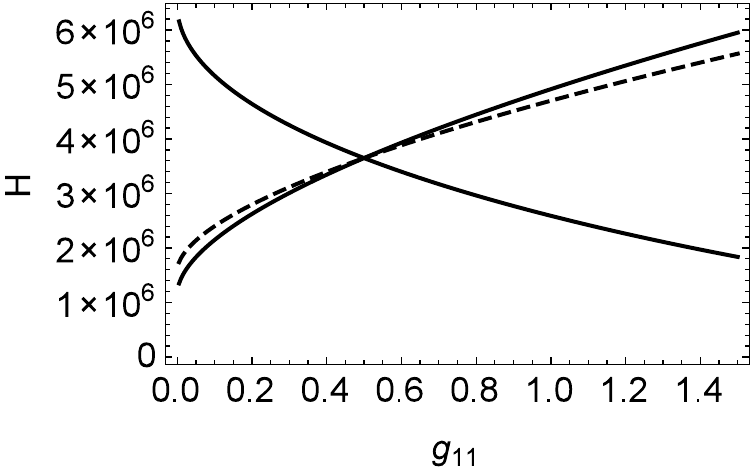} \label{MinEnergyA}}
        \subfigure[]{\includegraphics[width=0.47\linewidth]{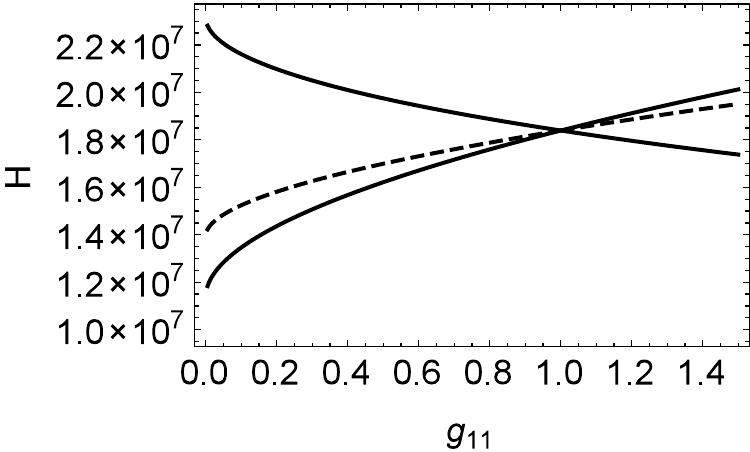} \label{MinEnergyB}}
        \caption{\textit{Comparison of the energies for the symmetric phases (solid lines) and the asymmetric immiscible phase (dashed line) as a function of the interaction constant $g_{11}$.
        For the symmetric case, the solid curves correspond to two configurations where the first component is inside, while the second one is outside and vice versa. In panel~\subref{MinEnergyA} the curves correspond to $g_{22}=0.5$ and the same number of particles in both components: $N_1:N_2=10000:10000$. In panel~\subref{MinEnergyB} we adopted $g_{22}=1$ and $N_1:N_2=10000:30000$. The particle masses and trap frequencies are identical: $m_1=m_2=1$ and $\omega_1=\omega_2=1$.}}
        \label{MinEnergy}
\end{figure}

The dash-dotted curve in Fig.~\ref{PhaseDiag} separates the diagram
into two regions: the region where the first component
is external, while the second one is internal
(the region above the dash-dotted curve), as shown in
Fig.~\ref{TomasFermi}\subref{TomasFermiA}-\subref{TomasFermiE}, and the region where the second component is external, while the first one is internal (the
region below the dash-dotted curve), as shown in (Fig.~\ref{TomasFermi}\subref{TomasFermiF}-\subref{TomasFermiJ}). Equating the coordinates where the densities
of the external and internal components become
zero, we will obtain the following relation for the miscible
components:
    \begin{eqnarray} \label{mw=mu}
    \frac{\mu_1}{\mu_2}=\frac{m_1\omega_1^2}{m_2\omega_2^2}.
    \end{eqnarray}
From this condition and Eqs. (\ref{NormRo}) we will derive the
following equation for the dash-dotted curve for the
region where $g_{12}^2<g_{11}g_{22}$:
    \begin{eqnarray} \label{}
    \frac{g_{12}}{g_{22}}=\frac{1}{\frac{N_1}{N_2}\frac{m_2\omega_2^2}{m_1\omega_1^2}\frac{g_{11}}{g_{12}}
    +\frac{m_2\omega_2^2}{m_1\omega_1^2}-\frac{N_1}{N_2}}.
    \end{eqnarray}
As we see, it is a hyperbola in the $(g_{11}/g_{12},g_{12}/g_{22})$ plane. In the case of immiscible components, comparison
of the energies for symmetric configurations
shows (see~Fig.~\ref{MinEnergy}) that the internal and external components also change places (see Figs.~\ref{TomasFermiE} and \ref{TomasFermiF}) when
passing through the hyperbola on which the equality
$g_{22} = g_{11}$ holds. Our numerical calculations show that
this hyperbola does not depend on the number of particles
in the components. This completes the construction
of a phase diagram in the Thomas-Fermi
approximation.
In general terms,

In general terms, the constructed diagram gives a
correct idea of the pattern of the component distributions
in traps, except for the region near the part of the
hyperbola that separates the distributions of types Fig.~\ref{TomasFermiE}
and \ref{TomasFermiF}. The point is that on this curve not only
the energies of the symmetric distributions in Fig.~\ref{TomasFermiE} and \ref{TomasFermiF} but also the energy of the asymmetric distribution,
where the components are on different sides of
the trap center (see~Fig.~\ref{TomasFermiK}), are equal to the same
value. As a result of such a degeneracy of the energies,
which is illustrated in Fig.~\ref{MinEnergy}, even a small perturbation
makes one of the distributions energetically more
favorable. As our numerical calculations show, allowance
for the dispersion gives an advantage to the asymmetric
phase in Fig.~\ref{TomasFermiK}. This difference is not captured
by the Thomas-Fermi approximation and requires a
more accurate calculation. The above classification of
the possible initial states that the condensate has
before the trap is switched off is sufficient for our purposes.

It should be noted that the distributions found have
breaks at the transition points from the overlap regions
to the singlet ones. Clearly, the dispersion effects at
these points also become significant and lead to a
smoothing of the curves. In particular, on the solid
curves the sharpening in the distribution at the center,
where the density of the second component in the
Thomas-Fermi approximation is zero, is smoothed
out, and a numerical solution of the Gross-Pitaevskii
equation gives a relatively small, but nonzero density at
the center (see, e.g.,~\cite{Quach-14}).

Fig.~\ref{PhaseDiag} shows a general phase diagram for identical
masses and trap frequencies. The number of particles
in the second component is twice that in the first
one. If we change the ratio of the trap frequencies and
masses, then the point of intersection between the perpendicular
straight lines will move along the diagonal $g_{11}/g_{12}=g_{12}/g_{22}$. For example, the point of intersection
will move upward as the parameters of the first component
increase and downward as the constants of the
second component increase. When changing the
number of particles, the point of intersection between
the straight lines will be stationary, but the dash-dotted
curve the passage through which interchanges the
internal and external components in the miscibility
region will change. In contrast, for immiscible components
the dash-dotted curve will remain
unchanged. The regions between the curves where the
external component becomes zero at the center of
symmetry and the diagonal will also change. In particular,
as the number of particles in the first component increases, the region where the second component is
expelled from the trap center will grow, while the
region where the first component becomes zero will be
reduced.

As a result, we have arrived at a complete classification
of the possible initial distributions and can now
turn to our main problem on the condensate expansion
after the trap has been switched off.

\section{A self-similar solution for the condensate expansion dynamics}

As was found in \cite{KaganSS-96,CastinDum-96,DalfovoMSP-97,KaganSS-97,Kamchatnov-04}, during the expansion of a
one-component condensate it can be assumed with a
good accuracy that the dependence of the density distribution
on the spatial coordinate does not change in
pattern, and the entire time dependence consists only
in the evolution of the parameters of this distribution
and the emergence of a distribution of the flow velocity
proportional to the coordinate. As a result, the
problem can be reduced to the solution of ordinary
differential equations for the distribution parameters,
and the solution can be found in a closed form in the
most interesting characteristic cases. In the two-component
case, this approach can have only a limited
applicability. For example, if the repulsion between
atoms in the internal component is much greater than
the interaction forces between atoms in the external
one, so that the initial distribution of the internal component
is shaped mainly by the trap potential, then
after the trap has been switched off, the pressure in the
internal component will be a dominant force and the
internal component will act on the external one like a
"piston". Nevertheless, if the difference between the
parameters of the two components is not too large,
then the time-evolving distributions will retain their
initial shape with a good accuracy during the expansion,
and, as in the one-component case, the problem
can be reduced to solving the equations for the distribution
parameters. The condition for this approximation
to be applicable is that each component evolves
predominantly under the action of its own pressure. In
addition, if the components are separated, then the
condition of mechanical equilibrium at the boundary
between them must be fulfilled. This means that
mechanical equilibrium is established in a time much
shorter than the characteristic expansion time until
the stage of motion by inertia, i.e., $R/c_s\ll1/\omega$, were $R$ is the characteristic size of the condensate, and $c_s=\sqrt{g\rho/m}$ is the sound velocity in the condensate component.
We will begin our discussion of the expansion
dynamics precisely with this case.

Thus, we will seek a time-dependent solution in a
form analogous to the initial distributions (\ref{IDfo}) and (\ref{IDfs}). More specifically, suppose that the density and
flow velocity depend on the coordinate quadratically and linearly, respectively, with time-dependent coefficients:
    \begin{eqnarray} \label{DA}
    \rho_i^n(x,t)=\alpha_{i,0}^n(t)-\alpha_i^n(t) x^2, \qquad
    u_i^n(x,t)=\beta_i^n(t) x,
    \end{eqnarray}
where $n=o,s$. Here, as before, the indices "o" and "s"
denote the quantities corresponding to the overlap and
singlet regions, respectively.

Substituting (\ref{DA}) into the continuity equation (\ref{CE1D}) and the Euler equation (\ref{EE1D}) gives
    \begin{eqnarray} \label{SoAlpha}
    -\dot\alpha_{i,0}^n=\alpha_{i,0}^n\beta_i^n, \qquad
    -\dot\alpha_i^n=3\alpha_i^n\beta_i^n,
    \end{eqnarray}
    \begin{equation} \label{SoBeta}
    \begin{split}
    &-\dot\beta_i^o=(\beta_i^o)^2-2\frac{g_{ii}}{m_i}\alpha_i^o-2\frac{g_{ij}}{m_i}\alpha_j^o+\omega_i(t)^2, \\
    &-\dot\beta_i^s=(\beta_i^s)^2-2\frac{g_{ii}}{m_i}\alpha_i^s+\omega_i(t)^2
    \end{split}
    \end{equation}
(the dot denotes a time derivative). Introducing a new
variable $\zeta_i^n$ defined by the relation
    \begin{equation} \label{al-zeta}
    \alpha_i^n=\frac{m_i}{2g_{ii}(\zeta_i^n)^3}.
    \end{equation}
we simplify considerably these equations: from (\ref{SoAlpha}) we
find that $\beta_i^n=\dot\zeta_i^n/\zeta_i^n$, and Eqs. (\ref{SoBeta}) will then take the
form
    \begin{equation} \label{EOMfo}
    \begin{split}
    \ddot\zeta_i^o=\frac{1}{(\zeta_i^o)^2}+\frac{m_j}{m_i}\frac{g_{ij}}{g_{ii}}\frac{\zeta_i^o}{(\zeta_j^o)^3}-\omega_i(t)^2\zeta_i^o,\qquad
    \ddot\zeta_i^s=\frac{1}{(\zeta_i^s)^2}-\omega_i(t)^2\zeta_i^s.
    \end{split}
    \end{equation}
This system of six second-order differential equations
defines the motion of the Bose-Einstein condensate
components. The last terms of the equations reflect
the influence of the confining potential on the
motion, while the other terms arise from the
interaction between atoms. In what follows, we will be
interested in the condensate dynamics after the trap
has been switched off, i.e., we should set $\omega_i(t)=0$ at $t>0$. The initial conditions for these equations are determined
by the original configurations that were
described in the previous section. At fixed nonlinear
constants they depend on the trap parameters and the
number of particles in each component.

    \begin{figure}[t] \centering
        \subfigure[]{\includegraphics[width=0.45\linewidth]{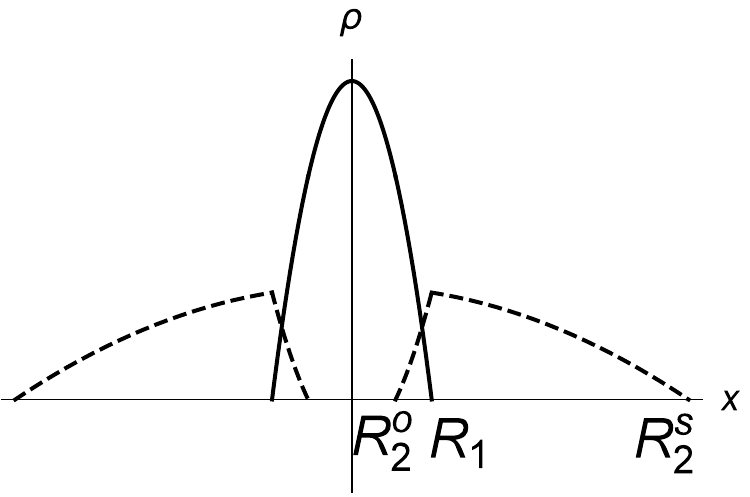} \label{GeneralA}}
        \subfigure[]{\includegraphics[width=0.45\linewidth]{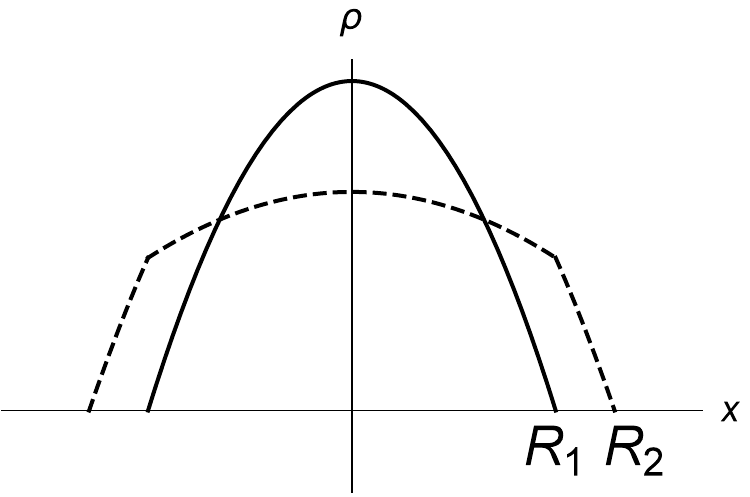} \label{GeneralB}}
        \caption{\textit{Two characteristic cases for the particle number densities: \subref{GeneralA} the first component (solid line) has a singlet region, \subref{GeneralB} the components are miscible everywhere. The components are numbered in such a way that the first component is always surrounded by the second one.}}
        \label{General}
    \end{figure}

Having solved the system of equations (\ref{EOMfo}), we can
find the velocities $u_i^n(x,t)$ and coefficients $\alpha_i^n(t)$. To find $\alpha_{i,0}^n(t)$, we will use the normalization of the wave
functions (\ref{NormRo}) and the fact that at the boundary
between the singlet and overlap regions the pressures
in them and, consequently, the densities are equal. To
be specific, we number the condensate components in
such a way that the first component is surrounded by
the second one, as shown in Fig.~\ref{General} (the first and second components are indicated by the solid and dashed
lines, respectively), so that we have
    \begin{equation} \label{Nofc}
    \begin{split}
    &\int_0^{R_2^o} \rho_1^odx+\int_{R_2^o}^{R_1} \rho_1^odx=\frac{1}{2}N_1, \\
    &\int_{R_2^o}^{R_1} \rho_2^odx+\int_{R_1}^{R_2^s} \rho_2^sdx=\frac{1}{2}N_2; \\
    &\rho_2^o(R_1,t)=\rho_2^s(R_1,t).
    \end{split}
    \end{equation}
Here $N_1$ and $N_2$ are are the numbers of particles in the
first and second components, respectively, $R_1$ is the
coordinate where the first component becomes zero ($\rho_1^o(R_1,t)=\alpha_{1,0}^o(t)-\alpha_1^o(t) R_1^2=0$), $R_2^o$ and $R_2^s$ are the
coordinates where the second condensate component
becomes zero ($\rho_2^n(R_2^n,t)=\alpha_{2,0}^n(t)-\alpha_2^n(t) {R_2^n}^2=0$). The point $R_2^o$ corresponds to zero density of the second
component in the overlap region, while $R_2^s$ denotes the coordinate at which the density of the second
component becomes zero in the singlet region (see~Fig.~\ref{GeneralA}). Consequently, these parameters are
defined by the relations
    \begin{equation}
    \begin{array}{ll}
    R_1=\sqrt{\frac{\alpha_{1,0}^o(t)}{\alpha_1^o(t)}},\qquad
    R_2^o=\sqrt{\frac{\alpha_{2,0}^o(t)}{\alpha_2^o(t)}},\qquad
    R_2^s=\sqrt{\frac{\alpha_{2,0}^s(t)}{\alpha_2^s(t)}}.
    \end{array}
    \end{equation}
The coordinates $R_1$, $R_2^o$ and $R_2^s$ are functions of time.
Thus, we have reduced the problem to integrating the
ordinary differential equations (\ref{EOMfo}) with their initial
conditions determined by the original component
density distributions in the trap. In general, this system
must be solved numerically, which is considerably easier
than the solution of the Gross-Pitaevskii equations.
However, an important case where system (\ref{EOMfo}) is simplified considerably and admits the derivation of
some relations in a closed form is noteworthy.

\subsection{The Case of Miscibility}

Let we have an initial overlap configuration where
the first component has no singlet region (see~Fig.~\ref{TomasFermiA}-\subref{TomasFermiC} and Fig.~\ref{TomasFermiH}-\subref{TomasFermiJ}), i.e., the components are miscible.
System (\ref{Nofc}) will then be written as
    \begin{equation}
    \begin{split}
    &\int_{0}^{R_1}\rho_1^odx=\frac{1}{2}N_1, \\
    &\int_0^{R_1}\rho_2^odx+\int_{R_1}^{R_2}\rho_2^sdx=\frac{1}{2}N_2;\\
    &\rho_2^o(R_1,t)=\rho_2^s(R_1,t).
    \end{split}
    \end{equation}
Here, $R_2$ is the coordinate where the particle number
density of the second component becomes zero (see~Fig.~\ref{GeneralB}). In this case, the relations for $\alpha_{i,0}^n(t)$ expressed
via $\alpha_{i}^n(t)$ can be found analytically, and the solution of
this system will be
    \begin{equation}
    \begin{split}
    &\alpha_{1,0}^o=\left[\frac{3}{4}N_1\sqrt{\alpha_1^o}\right]^{2/3}, \\
    &\alpha_{2,0}^s=\left[\frac{3\sqrt{\alpha_2^s}}{4\alpha_1^o}(N_2\alpha_1^o-N_1(\alpha_2^o-\alpha_2^s))\right]^{2/3}, \\
    &\alpha_{2,0}^o=\alpha_{2,0}^s+\left[\frac{3N_1}{4\alpha_1^o}\right]^{2/3} (\alpha_2^o-\alpha_2^s).
    \end{split}
    \end{equation}
The case of the expansion of a condensate with an
asymmetric initial profile (see~Fig.~\ref{TomasFermiK}) will be considered
separately below.

    \begin{figure}[t] \centering
        \subfigure[]{\includegraphics[width=0.45\linewidth]{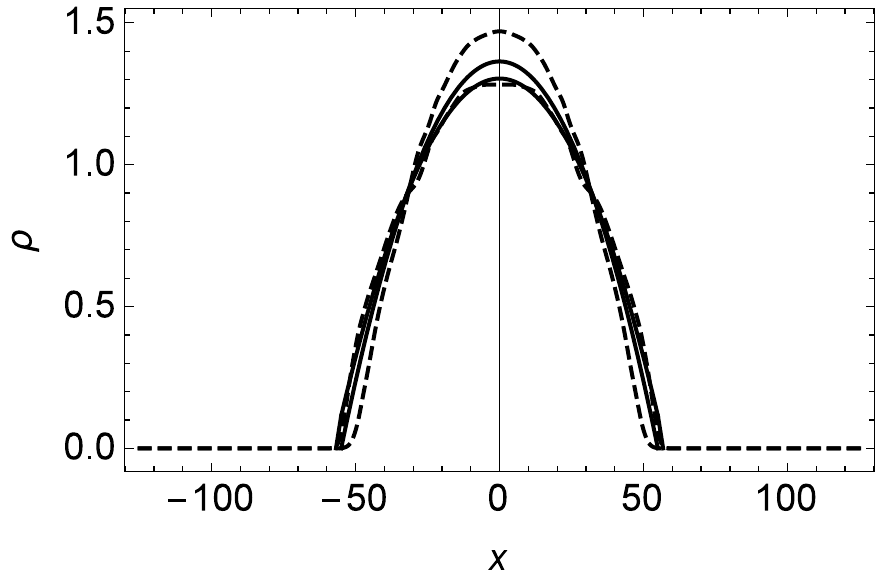} \label{NumericalA}}
        \subfigure[]{\includegraphics[width=0.45\linewidth]{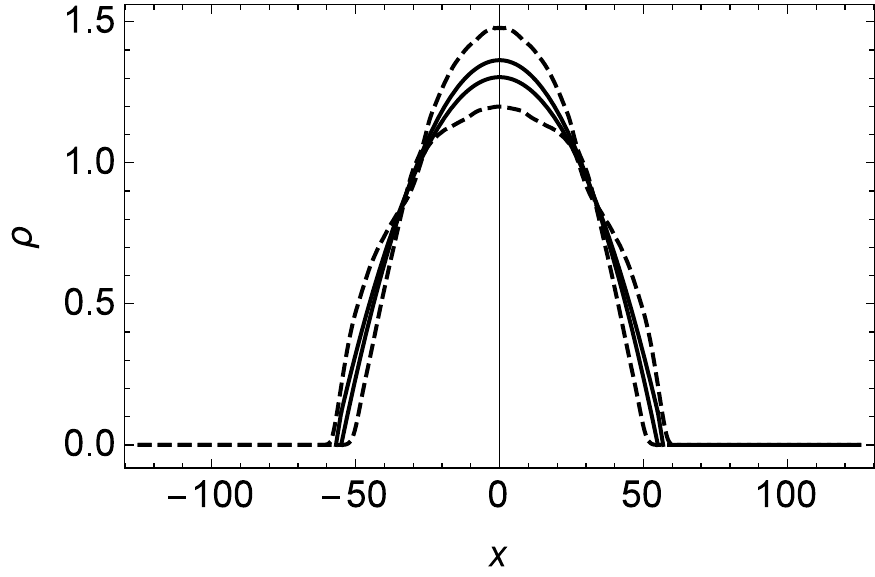} \label{NumericalB}}
        \caption{\textit{Comparison of the numerical solution of the Gross-Pitaevskii equations (\ref{GPEtcod}) (dashed curves) with the elf-similar solution of the equations of motion (\ref{EOMfo}) (solid curves) for $g_{11}=3$, $g_{22}=4$, $g_{12}=0.1$ \subref{NumericalA} and $g_{11}=3$, $g_{22}=4$, $g_{12}=2$ \subref{NumericalB} with the same number of particles in the components $N_1=N_2=100$ at time $t=5$. The particle masses and trap frequencies are $m_1=m_2=1$ and $\omega_1=\omega_2=1$.}}
        \label{Numerical}
    \end{figure}

The differential equations (\ref{EOMfo}) are Newton-type
equations that have the total energy of the system (\ref{Hamil}) the conservation of the number of particles in the
singlet region and the overlap region of each of the components as the integrals of motion. Generally,
these integrals are not enough to find the analytical
solution of the system. Therefore, we turn to its
numerical solution. Figure~\ref{Numerical} shows examples of comparing
the numerical solution of the Gross-Pitaevskii
equations (\ref{GPEtcod}) (dashed curves) with the numerical
solution of the equations of motion (\ref{EOMfo}) (solid curves)
for different interaction constants at fixed times. As
can be seen from the figure, in the region where the
components are well miscible the particle density distributions
retain their shape during the expansion, and
the self-similar solution quantitatively describes the
dynamics of the system excellently. A change in the
nonlinear interaction constant between the components
by a factor of $10$ does not affect significantly the
accuracy of the approximation as long as the miscibility
criterion (\ref{gg}) holds with a margin (in the case of Fig.~\ref{NumericalB}, $g_{12}^2=4$ is smaller than $g_{11}g_{22}=12$ by a factor
of $3$).

\begin{figure}[t] \centering
        \includegraphics[width=0.45\linewidth]{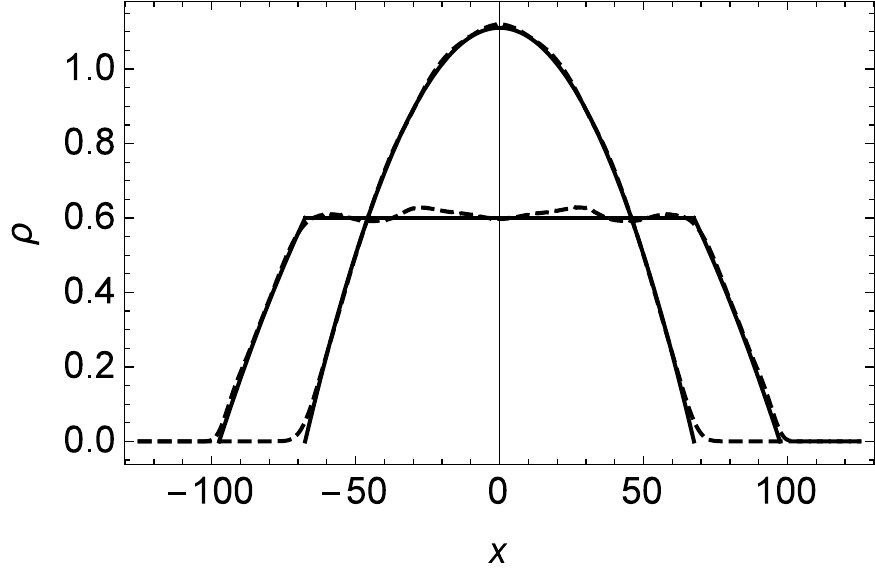}
        \caption{\textit{Comparison of the numerical solution of the Gross-Pitaevskii equations (\ref{GPEtcod}) (dashed curve) with the numerical solution of Eqs. (\ref{xig=g1}), (\ref{xig=g2}) (solid curve) for $g_{11}=g_{12}=1$, $g_{22}=2$ with the same number of particles in the components $N_1:N_2=100:100$ at time $t=10$. The particle masses and trap frequencies are $m_1=m_2=1$ and $\omega_1=\omega_2=1$.}}
        \label{g=gEOM}
    \end{figure}

In practice the interaction constants have almost
the same value in many cases. For example, for an $^{87}$Rb atom in different states of the hyperfine structure ($|1,-1\rangle$ and $|2,-2\rangle$) the scattering lengths are $a_{11}=98.98a_0$, $a_{12}=98.98a_0$ and $a_{22}=100.4a_0$, where $a_0$ the Bohr radius (see, e.g.,~\cite{VerhaarKvK-09,Kamchatnov-13}) i.e., the interaction
constants are also equal, $g_{11}=g_{12}$. The component
masses and trap frequencies can often be also equal ($m_1=m_2$, $\omega_1=\omega_2$). The second (external) component
of the Bose-Einstein condensate will then be constant
in the overlap region and, consequently, $\alpha_2^o(t)=0$. Consider a slightly more general case where the condition (\ref{polochka}) is fulfilled. The self-similar solution will then
be written as
    \begin{equation}
    \begin{array}{l}
    \rho_1^o(x,t)=\alpha_{1,0}^o(t)-\alpha_1^o(t) x^2,\qquad
    \rho_2^o(x,t)=\alpha_{2,0}^o(t),\qquad
     \rho_2^s(x,t)=\alpha_{2,0}^s(t)-\alpha_2^s(t) x^2;\\
    u_i^n(x,t)=\beta_i^n(t)x.
    \end{array}
    \end{equation}
After the substitution into Eqs. (\ref{CE}) and (\ref{EE}) the self-similar
solution will give equations analogous to (\ref{SoAlpha}) and (\ref{SoBeta}):
    \begin{equation} \label{SoBetag}
     \begin{array}{l}
    -\dot\alpha_{1,0}^o=\alpha_{1,0}^o\beta_1^o,\quad
    -\dot\alpha_{2,0}^o=\alpha_{2,0}^o\beta_2^o,\quad
    -\dot\alpha_{2,0}^s=\alpha_{2,0}^s\beta_2^s;\qquad
    -\dot\alpha_1^o=3\alpha_1^o\beta_1^o,\quad
    -\dot\alpha_2^s=3\alpha_2^s\beta_2^s;\\
    -\dot\beta_1^o=(\beta_1^o)^2-2\frac{g_{11}}{m_1}\alpha_1^o,\qquad
    -\dot\beta_2^o=(\beta_2^o)^2-2\frac{g_{12}}{m_2}\alpha_1^o,\qquad
    -\dot\beta_2^s=(\beta_2^s)^2-2\frac{g_{22}}{m_2}\alpha_2^s.
    \end{array}
    \end{equation}
By substituting $\alpha_1^o={m_1}/{2g_{11}(\zeta_1^o)^3}$ and $\alpha_2^s={m_2}/{2g_{22}(\zeta_2^s)^3}$, the system of equations at $\omega_i(t)=0$ is reduced to
    \begin{eqnarray} \label{EOMfg}
    \begin{array}{ll}
    \ddot\zeta_1^o=\frac{1}{(\zeta_1^o)^2},\qquad
    \ddot\zeta_2^s=\frac{1}{(\zeta_2^s)^2}.
    \end{array}
    \end{eqnarray}
From the first equation (\ref{EOMfg}) we will find
    \begin{equation} \label{xig=g1}
    \begin{split}
    \sqrt{2} \Omega t=\Omega^{1/3}\sqrt{\zeta_1^o(\Omega^{2/3}\zeta_1^o-1)} +\frac{1}{2}\ln({2\Omega^{2/3}\zeta_1^o+2\Omega^{1/3}\sqrt{\zeta_1^o(\Omega^{2/3}\zeta_1^o-1)}-1})
    \end{split}
    \end{equation}
where we introduce an effective frequency of the
potential in which the first condensate component is
located:
    $$
    \Omega=\left(\frac{g_{11}}{m_1}\frac{g_{22}m_1\omega_1(0)^2-g_{12}m_2\omega_2(0)^2}{g_{11}g_{22}-g_{12}^2} \right)^{1/2}.
    $$
While from the second equation (\ref{EOMfg}) we obtain
    \begin{equation} \label{xig=g2}
    \begin{split}
    \sqrt{2}\omega_2 t=\omega_2^{1/3}\sqrt{\zeta_2^s(\omega_2^{2/3}\zeta_2^s-1)} +\frac{1}{2}\ln({2\omega_2^{2/3}\zeta_2^s+2\omega_2^{1/3}\sqrt{\zeta_2^s(\omega_2^{2/3}\zeta_2^s-1)}-1}).
    \end{split}
    \end{equation}
Equations (\ref{xig=g1}) and (\ref{xig=g2}) implicitly specify $\zeta_i^n$ as a function of $t$. The densities derived from these equations
and by numerically solving the Gross-Pitaevskii
equations are compared in Fig.~\ref{g=gEOM}.

The velocities expressed via $\zeta_i^n$ can also be easily
found from system (\ref{EOMfg}):
    \begin{equation}
    \begin{split}
     u_1^o=\frac{\sqrt{2}x}{\zeta_1^o}\sqrt{\Omega^{2/3}-\frac{1}{\zeta_1^o}}, \qquad
     u_2^s=\frac{\sqrt{2}x}{\zeta_2^s}\sqrt{\omega_2^{2/3}-\frac{1}{\zeta_2^s}}.
     \end{split}
    \end{equation}
At $t\gg\omega_i^{-1}$ from (\ref{xig=g1}) and (\ref{xig=g2}) we find the asymptotic formulas
    \begin{eqnarray} \label{xiog=g}
    \zeta_1^o\approx\sqrt{2}\Omega^{1/3}t, \qquad
    \zeta_2^s\approx\sqrt{2}\omega_2^{1/3}t.
    \end{eqnarray}
These solutions correspond to the motion of condensate
atoms by inertia, when the density in the condensate
cloud becomes so low that the pressure no longer
accelerates the condensate.

To find the velocity of the second component in
the overlap region, we will use (\ref{xiog=g}) and the second equation (\ref{SoBetag}) and find a differential equation for $\beta_2^o$
    \begin{equation}
    -\dot\beta_2^o\approx(\beta_2^o)^2-\frac{m_1}{m_2}\frac{g_{12}}{g_{11}}\frac{1}{(\sqrt{2}t)^3\Omega}.
    \end{equation}
From this equation we will obtain
    \begin{equation}
    \beta_2^o\sim\frac{1}{t}, \qquad t\gg\omega_i^{-1}.
    \end{equation}
Knowing the asymptotic solutions for $\zeta_i$ and the velocities
expressed via these $\zeta_i$ we can easily find the
asymptotic solution for the velocities of each of the
components $u_i^n$:
    \begin{equation}
    u_i^n(x,t)\sim\frac{x}{t}.
    \end{equation}
The extreme points of the distribution of each condensate
component will move with the greatest velocities:
    \begin{equation} \label{VelosFg=g1}
    \begin{split}
    &u_{1,max}^o\approx\left(3\sqrt{2}N_1\frac{g_{11}\Omega}{m_1}\right)^{1/3}, \\
    &u_{2,max}^s\approx\left(3\sqrt{2}(N_2\frac{g_{22}\omega}{m_2}+N_1\frac{g_{11}\Omega}{m_1})\right)^{1/3}.
    \end{split}
    \end{equation}
The first formula (\ref{VelosFg=g1}) corresponds to the velocity of
the distribution boundary for the first condensate
component, when the motion occurs by inertia, while
the second formula represents the same velocity for
the second component. From the asymptotics found
for $\zeta_i$ we can also derive simple formulas for the density
distributions at $t\gg\omega_i^{-1}$:
    \begin{eqnarray}
    && \rho_1^o(x,t)\approx\frac{m_1}{4\sqrt{2}g_{11}\Omega}{{u_1^o}^2_{max}}\frac{1}{t} \left(1-\frac{x^2}{({{u_1^o}_{max}}t)^2}\right), \\
    && \rho_2^o(x,t)\approx\frac{m_2}{4\sqrt{2}g_{22}\omega_2}({{u_2^s}^2_{max}}-{{u_1^o}^2_{max}})\frac{1}{t}, \\
    && \rho_2^s(x,t)\approx\frac{m_2}{4\sqrt{2}g_{22}\omega_2}{{u_2^s}^2_{max}}\frac{1}{t} \left(1-\frac{x^2}{({{{u_2^s}_{max}}}t)^2}\right).
    \end{eqnarray}

Thus, in the case of strong miscibility of the condensate
components with initial distributions like
those in Fig.~\ref{TomasFermiA},\subref{TomasFermiB},\subref{TomasFermiI},\subref{TomasFermiJ}, the self-similar solution
gives quite a satisfactory description of the condensate
expansion after the trap has been switched off.
If, however, we approach the miscibility-immiscibility
boundary with initial distributions like those in Fig.~\ref{TomasFermiC},\subref{TomasFermiD},\subref{TomasFermiG},\subref{TomasFermiH}, then the Thomas-Fermi
approximation loses its accuracy even when calculating
the stationary distributions due to the appearance
of large jumps in derivatives in the density distributions.
The expansion dynamics in such cases also differs
significantly from the predictions of the self-similar
theory. In particular, characteristic regions of nonlinear
oscillations can be formed in the density and
velocity distributions during the evolution. This means
that wave breaking occurs under their deformation due
to nonlinear effects, so that the dispersion effects can
no longer be neglected. In contrast, allowance for the
simultaneously nonlinear and dispersion effects gives
rise to a region of oscillations connecting the regions
of flows with different parameters. These regions of
oscillations are analogous to shock waves in low-dissipation
systems, and they were called "dispersive shock
waves". However, before turning to their study, let us
consider the case where the components are strongly immiscible, which is also described by a self-similar
solution analogous to those obtained above for miscible
components.

\subsection{The Case of Immiscible Components}

Here, we assume that the Bose-Einstein condensate
components are absolutely immiscible, i.e., the
overlap region is of the order of the correlation length.
As has been pointed out in Section \ref{SectionPD}, symmetric (Fig.~\ref{TomasFermiE},\subref{TomasFermiF}) distributions
can be realized in this case. The self-similar solution
will now take the form
    \begin{eqnarray} \label{eq:al-be}
    \rho_i^s(x,t)=\alpha_{i,0}^s(t)-\alpha_i^s(t)x^2, \qquad
    u_i^s(x,t)=\beta_i^s(t)x
    \end{eqnarray}
and out of Eqs. (\ref{EOMfo}) only those responsible for the singlet
regions will remain. The initial distributions $\alpha_{i,0}^s$ and $\alpha_{i}^s$ are defined by the stationary solutions
    \begin{eqnarray}
    &&\rho_i^s(x,0)=|\psi_i^s|^2=\frac{2\mu_i-m_i\omega_i^2x^2}{2g_{ii}}.
    \end{eqnarray}
At the boundary between the immiscible components
the pressures in them are equal, $p_i=p_j$, which serves as
a condition defining the boundary coordinate. In the
hydrodynamic approximation the pressures are $p_i=g_{ii}{\rho_i^s}^2/2m_i$. So, from this condition we find that the
coordinate $R_{b}$ of the boundary between the components
is given by the formula
    \begin{equation}
    R_{b}=\left(\frac{\sqrt{g_{11}m_2}\alpha_{1,0}^s-\sqrt{g_{22}m_1}\alpha_{2,0}^s}
    {\sqrt{g_{11}m_2}\alpha_{1}^s-\sqrt{g_{22}m_1}\alpha_{2}^s}\right)^{1/2}.
    \end{equation}
Thus, the problem has been reduced to determining
six functions of time that are the coefficients in (\ref{eq:al-be}).

The expressions for $\alpha_{i}^s(t)$ can be found in a closed
form. At $t>0$, where $\omega_i(t)=0$, we will obtain simple
equations of motion for the parameters of the singlet
distributions:
    \begin{equation} \label{EOMfim}
    \ddot\zeta_i^s=\frac{1}{(\zeta_i^s)^2}, \qquad i=1,2
    \end{equation}
($\alpha_i^s=m_i/2g_{ii}(\zeta_i^s)^3$). It is easy to find the solution of this
system:
    \begin{equation} \label{xiIm}
    \begin{split}
    \sqrt{2}\omega_i t=\omega_i^{1/3}\sqrt{\zeta_i^s(\omega_i^{2/3}\zeta_i^s-1)}+\frac{1}{2}\ln({2\omega_i^{2/3}\zeta_i^s+2\omega_i^{1/3}\sqrt{\zeta_i^s(\omega_i^{2/3}\zeta_i^s-1)}-1}).
    \end{split}
    \end{equation}
These equations implicitly specify $\zeta_i^s$ as functions of $t$. The functions $\beta_i^s(t)={\dot\zeta_i^s}/{\zeta_i^s}$ are expressed via the
functions $\zeta_i^s$ found. As a result, we obtain the expressions
for the velocity of each component $u_i^s$ expressed
via $\zeta_i^s$:
    \begin{equation}
    u_i^s=\beta_i^sx=\frac{\dot\zeta_i^s}{\zeta_i^s}x=\frac{\sqrt{2}x}{\zeta_i^s}\sqrt{\omega_i^{2/3}-\frac{1}{\zeta_i^s}}.
    \end{equation}
Finally, the relation between the parameters $\alpha_{i,0}^s$ and $\alpha_i^s$ follows from the normalization conditions (here as before $R_1=\sqrt{\alpha_{1,0}^s/\alpha_1^s}$ and $R_2=\sqrt{\alpha_{2,0}^s/\alpha_2^s}$):
    \begin{eqnarray}
    \label{NormImmis1s}
    \int_{0}^{R_b}\rho_1^sdx=\frac{N_1}{2},\qquad
    \label{NormImmis2s}
    \int_{R_b}^{R_2}\rho_2^sdx=\frac{N_2}{2}
    \end{eqnarray}
for the symmetric profile and from the equations
    \begin{eqnarray}
    \label{NormImmis1as}
    \int_{R_1}^{R_b}\rho_1^sdx=N_1,\qquad
    \label{NormImmis2as}
    \int_{R_b}^{R_2}\rho_2^sdx=N_2
    \end{eqnarray}
for the asymmetric one. In principle, these equations
allow $\alpha_{i,0}^s$ to be expressed via the already known functions $\alpha_{i}^s$ through numerically easily solvable algebraic
equations (which we do not write out here, because
they are cumbersome).

    \begin{figure}[t] \centering
        \subfigure[]{\includegraphics[width=0.45\linewidth]{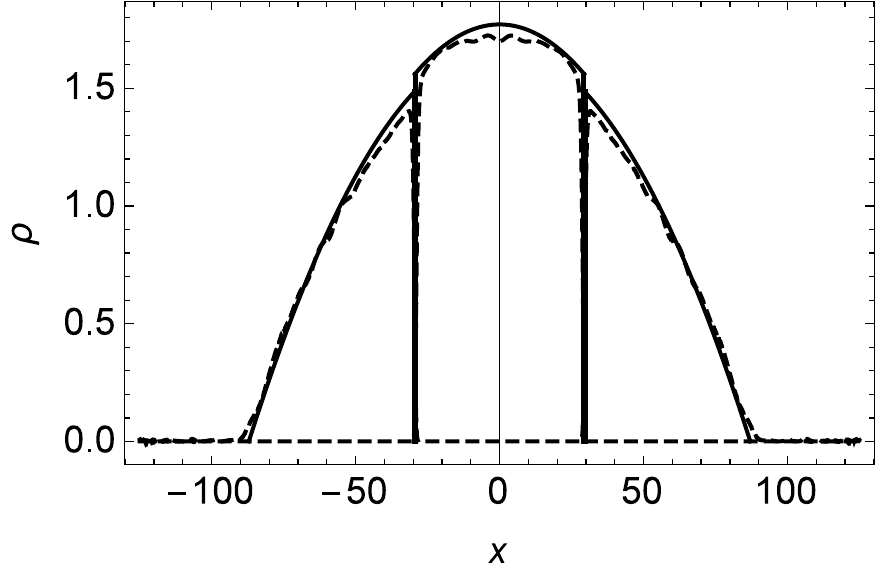} \label{ImmisA}}
        \subfigure[]{\includegraphics[width=0.45\linewidth]{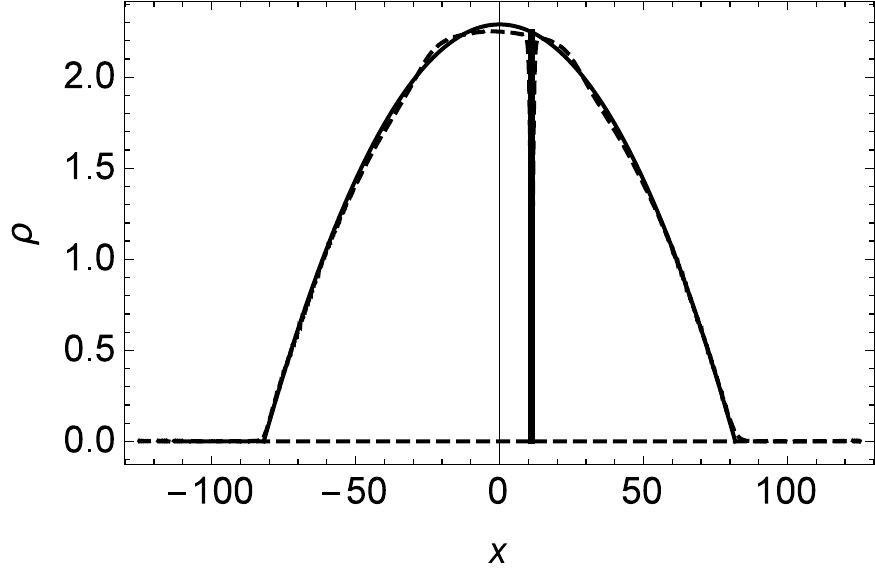} \label{ImmisB}}
        \caption{\textit{Particle number densities for the symmetric phase with $g_{11}=1$, $g_{22}=1.1$, and the same number of particles in the components $N_1:N_2=100:100$ \subref{ImmisA} and the asymmetric phase with $g_{11}=1$, $g_{22}=1$ and $N_1:N_2=150:100$ \subref{ImmisB} at $t=10$. The particle masses and trap frequencies are $m_1=m_2=1$, $\omega_1=\omega_2=1$.}}
        \label{Immis}
    \end{figure}

As in the case of $g_{11}=g_{12}$, from the system of equations (\ref{xiIm}) we can find the asymptotic solution for $\zeta_i^s$ corresponding to times $t\gg\omega_i^{-1}$:
    \begin{equation}
    \zeta_i^s\approx\sqrt{2}\omega_i^{1/3}t.
    \end{equation}
As above, they describe the flow of the condensate
cloud by inertia once the potential energy of the condensate
compressed in the trap has been converted
into the kinetic energy of its flow. The extreme points
of the distributions of each condensate component
move with maximum velocities:
    \begin{eqnarray}
    \label{VelosFg=gIms}
    &&{u_1^s}_{max}\approx\frac{R_b(t)}{t},\qquad
      {u_2^s}_{max}\approx\frac{R_2(t)}{t}; \\
    \label{VelosFg=gImas}
    &&{u_1^s}_{max}\approx\frac{R_1(t)}{t},\qquad
      {u_2^s}_{max}\approx\frac{R_2(t)}{t}
    \end{eqnarray}
for the symmetric and asymmetric particle number
density profiles, respectively. The expressions for the
particle number densities and velocities at asymptotically
large $t$ are
    \begin{eqnarray}
    \rho_i^s(x,t)\approx\alpha_{i,0}^s(t)-\frac{m_i}{4\sqrt{2}g_{ii}\omega_i}\frac{x^2}{t^3}, \qquad
    u_i^s(x,t)\sim\frac{x}{t}.
    \end{eqnarray}
The functions $\alpha_{i,0}^s(t)$ The functions (\ref{NormImmis2s}) in
the symmetric case and (\ref{NormImmis1as}) in the asymmetric case
for each time $t$. Knowing $\alpha_{1,0}^s$ and $\alpha_{2,0}^s$, we can find the
forms of the particle number densities and the maximum
velocities of each of the condensate components
at this time.

Thus, the derived formulas give the solution of our
problem. It is illustrated in Fig.~\ref{Immis}, where the particle
number densities are shown for the symmetric and
asymmetric phases at a fixed time.

\section{Dispersive shock wave} \label{SectionDSW}

\begin{figure}[t] \centering
        \subfigure[]{\includegraphics[width=0.45\linewidth]{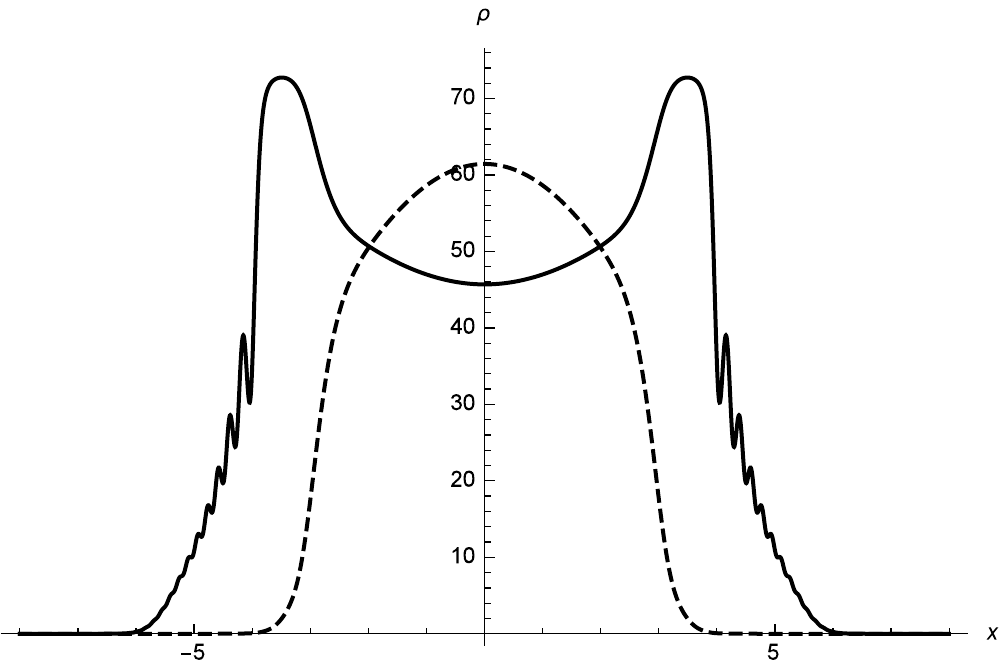} \label{ExampleShockwavesA}}
        \subfigure[]{\includegraphics[width=0.45\linewidth]{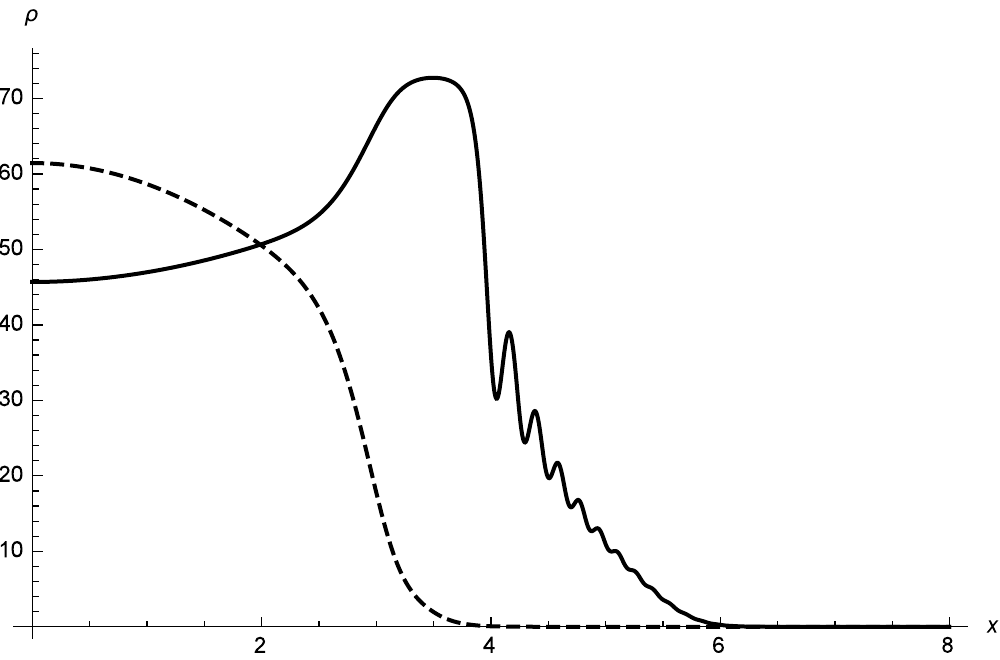} \label{ExampleShockwavesB}}
        \caption{\textit{Example of a shock wave for miscible components.}}
        \label{ExampleShockwaves}
    \end{figure}

A characteristic feature of the expansion of a twocomponent
Bose-Einstein condensate is the possibility
of the appearance of dispersive shock waves in it
(see~\cite{KamchatnovGK-04,HoeferACCES-06}). Fig.~\ref{ExampleShockwaves} shows an example of such
waves, where the internal component expels the external
one so strongly that this leads to a breaking of the
particle number density distribution of the external component,
which implies the beginning of the formation of a dispersive
shock wave. In the case where the external
component is much larger than shock wave size, we
can roughly assume that in the place of shock wave
formation the external component is homogeneous,
while the boundary of the internal component is a piston
pushing the external component. Similarly, the
formation of dispersive shock wave is also possible for
immiscible components. For example, in a condensate
confined in a trap a boundary is formed between the condensates at which the pressures of both components
are equal. If the frequencies of the traps confining
the various components differ greatly, then a pressure
jump arises at the boundary between the condensates
immediately after the traps have been switched
off, so that one component will again act on the other
component as a piston. At the initial stage the region
near the boundary between the condensates is of
greatest interest, while the condensate parameters far
from the boundary may be deemed to be specified by
constants. Such problems on a piston both with a constant
velocity of its motion and with a uniformly accelerated
or arbitrary law of its motion were considered in \cite{HoeferAE-08,KamchatnovKorneev-10} and can be applied to the problem on the condensate
expansion after the traps have been switched
off.

\begin{figure}[t] \centering
        \includegraphics[width=0.6\linewidth]{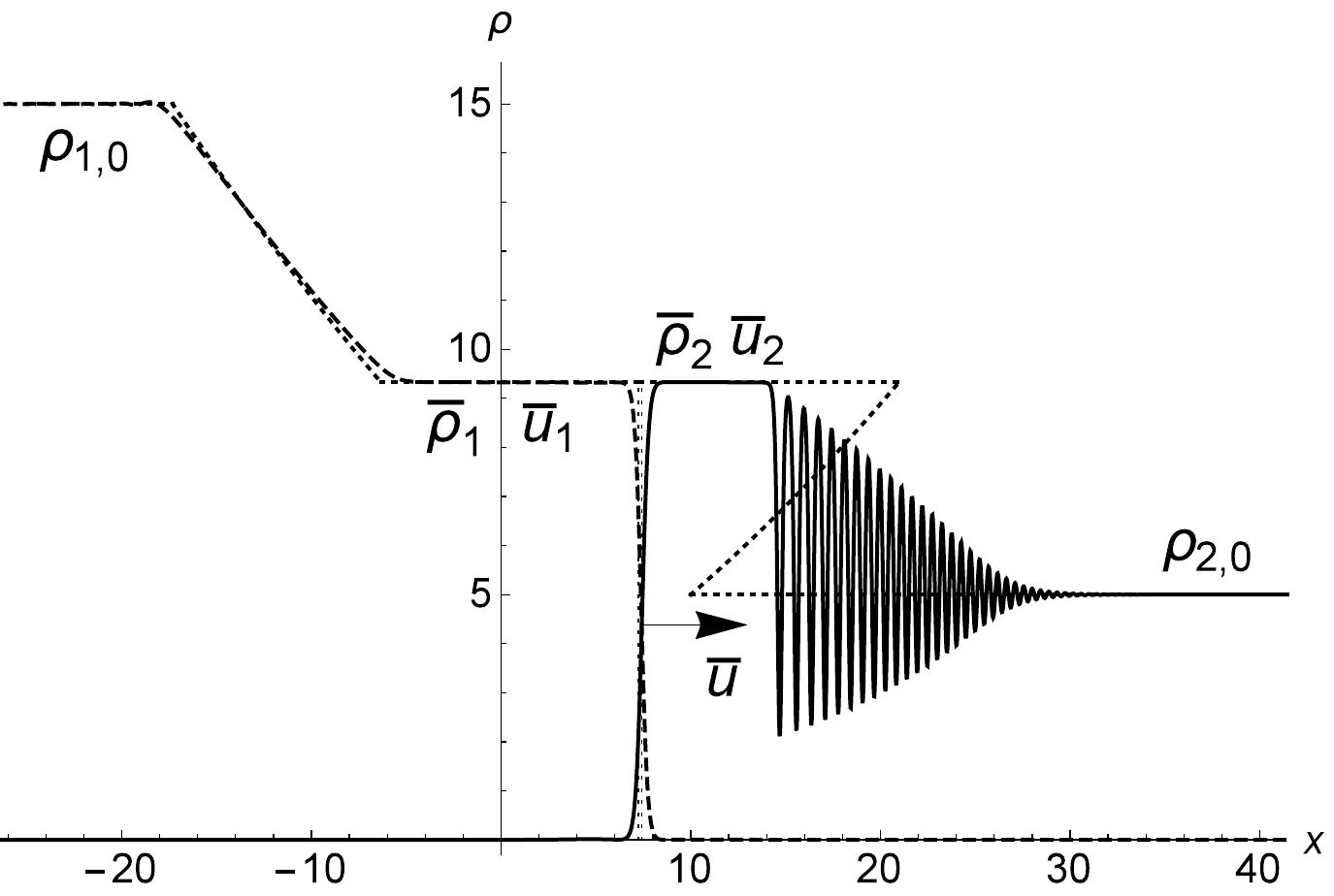}
        \caption{\textit{Numerical solution of the evolution of a step for immiscible components at $t=2$. The interaction constants are $g_{11}=5$, $g_{22}=5$, $g_{12}=5.5$. The initial densities are $\rho_{1,0}=15$ and $\rho_{2,0}=5$. The dashed and solid lines indicate the left and right components, respectively; the dotted line represents the dispersionless limit.}}
        \label{Step}
    \end{figure}

As an example, consider an immiscible condensate
at an evolutionary stage when the size of the region
occupied by the dispersive shock wave is much smaller
than the size of the entire condensate cloud. We will
also assume that the internal component plays the role
of a piston. The evolution of the external component is
then described by a solution analogous to those found
in \cite{HoeferAE-08,KamchatnovKorneev-10}, but with allowance made for the fact that
now the piston velocity is not specified but must be
found in a self-consistent way from the condition that the pressures of the two components at the boundary
between them are equal. This means that for the internal
component we are also dealing with the piston
problem, but, in this case, the piston "stretches" the
condensate rather than compresses it. As a result, a
rarefaction wave described by the well-known solution
of the hydrodynamic equations (\ref{CE1D}) and (\ref{EE1D}). for the
internal component will propagate into the internal
condensate. A characteristic density profile of the
emerging wave structure is shown in Fig.~\ref{Step} it was
obtained by numerically solving the Gross-Pitaevskii
equations (\ref{GPEtcod}). Our task in this section is to calculate the
main characteristics of this structure by expressing
them via the condensate parameters and initial conditions.

Thus, we will assume that at the initial time, immediately
after the traps have been switched off, the first
component is located to the left of the interface at $x=0$ and has a density $\rho=\rho_{1,0}$, while the second component
is located to the right of the interface and has a
density $\rho=\rho_{2,0}$. The flow velocities of both components
are zero, $u_{1}=u_{2}=0$, at the initial time $t=0$. To
be specific, we will also assume that the pressure to the
left of the interface at the initial time is greater than the
pressure to the right of the interface:
    \begin{equation} \label{pressure-cond}
    p_1=\frac12g_{11}\rho_{1,0}^2>p_2=\frac12g_{22}\rho_{2,0}^2.
    \end{equation}
Precisely such a condition corresponds to the formation
of the wave structure shown in Fig.~\ref{Step}. Clearly, this
condition does not limit the generality of our consideration:
if the inequality is opposite, then the waves
will propagate in the opposite direction, and it will be
sufficient to replace $x$ by $-x$ in the formulas derived
below in order that they describe this case.

In accordance with Fig.~\ref{Step} it follows from the theory
of compressible fluid motion \cite{LandauLifshitz-59} that the left rarefaction
wave connects two regions of homogeneous
flow: the condensate at rest with a density $\rho_{1,0}$ and the
``shel'' to the right of the rarefaction wave with some
density $\overline{\rho}_{1}$ and flow velocity $\overline{u}_{1}$. This shelf extends to
the boundary with the second condensate that moves
with a velocity $\overline{u}$ equal to the flow velocity $\overline{u}_{1}$ of the left
condensate. As has been pointed out above, the
boundary between the components serves as a ``piston''
for the second condensate. Therefore, in accordance
with the well-known solution \cite{HoeferAE-08}, a shelf of the
flow of the second condensate with a density $\overline{\rho}_{2}$ and
flow velocity $\overline{u}_{2}$ again equal to the boundary velocity $\overline{u}$ also emerges rightward of the boundary. This second
shelf is connected by the dispersive shock wave with
the region of the second component at rest with a density $\rho_{2,0}$. The boundary between the two condensates
is an analog of the contact discontinuity known from
the classification of the decays of initial discontinuities
in gas dynamics (see.~\cite{LandauLifshitz-59}). Owing to the equality of the
nonlinear constants, $g_{11}=g_{22}$ the equality of the densities follows from the equality of the pressures at such
a discontinuity. For simplicity, precisely such a case is
assumed in Fig.~\ref{Step} The parameters of the shelfs $\overline{\rho}_{1}$, $\overline{\rho}_{2}$, the velocity of the boundary (contact discontinuity) $\overline{u}=\overline{u}_{1}=\overline{u}_{2}$, and the velocities of the rarefaction
wave and dispersive shock wave boundaries are
required to be found in the problem formulated in this
way.

Let us first discuss this problem in terms of the dispersionless
hydrodynamic equations
    \begin{equation} \label{hydro}
    \begin{split}
    &\frac{\partial\rho_i}{\partial t}+\frac{\partial}{\partial x}(\rho_iu_i)=0,\\
    &\frac{\partial u_i}{\partial t}+u_i\frac{\partial u_i}{\partial x}+ g_{ii}\frac{\partial\rho_i }{\partial x}=0.
    \end{split}
    \end{equation}
These are a special case of Eqs. (\ref{CE1D}) and (\ref{EE1D}), where
we take into account the fact that, in this case, the
components are spatially separated, so that the index $i=1$ refers to the component to the left of the interface,
while the index $i=2$ refers to the component to the
right of this interface. There is no interaction between
the components everywhere except the narrow region
near the boundary separating them, but its presence is
taken into account by an appropriate boundary condition
for the pressures to be equal on both sides of the
boundary. In addition, we assume the masses of the
atoms in both components to be identical and equal to
unity, while the trap potential at the expansion stage to
play no role in an obvious way. The solution of such
problems is simplified considerably if we pass from the
ordinary physical variables $\rho_i$, $u_i$ to the so-called
``Riemann invariants''. For Eqs. (\ref{hydro}), which coincide
in form with the gas-dynamic equations, the Riemann invariants are well known and can be written as (see,
e.g.,~\cite{Kamchatnov-2000})
    \begin{eqnarray} \label{rim1}
    r_{\pm}^{(i)}=u_i\pm2\sqrt{g_{ii}\rho_i}.
    \end{eqnarray}
In these variables the hydrodynamic equations (\ref{hydro}) take a simple symmetric form:
    \begin{equation} \label{rim2}
    \frac{\prt r_{\pm}^{(i)}}{\prt t}+v_{\pm}^{(i)}(r_{+}^{(i)},r_{-}^{(i)})\frac{\prt r_{\pm}^{(i)}}{\prt x}=0,
    \end{equation}
where the ''Riemann velocities" $v_{\pm}^{(i)}=u_i\pm c_i$ are
expressed via the Riemann invariants by the relations
    \begin{equation} \label{rim3}
    v_+^{(i)}=\frac34r_{+}^{(i)}+\frac14r_{-}^{(i)},\qquad  v_-^{(i)}=\frac14r_{+}^{(i)}+\frac34r_{-}^{(i)}.
    \end{equation}
If the solution of Eqs. (\ref{rim2}) has been found and the
Riemann invariants are known, then the physical variables
are expressed via them by the formulas
    \begin{eqnarray} \label{raref}
    \rho_i=\frac{(r_{+}^{(i)}-r_{-}^{(i)})^2}{16g_{ii}},\qquad u_i=\frac12(r_{+}^{(i)}+r_{-}^{(i)}).
    \end{eqnarray}
A rarefaction wave is known \cite{LandauLifshitz-59} to belong to the
class of ``simple waves'' characterized by the fact that
one of the Riemann invariants has a constant value
along the flow. For the case shown in Fig.~\ref{Step}, the rarefaction
wave propagates to the left in the first condensate.
Hence, the following invariant is constant
for it:
    \begin{eqnarray} \label{r1-pl}
    r_{+}^{(1)}=u_1+2\sqrt{g_{11}\rho_1}=2\sqrt{g_{11}\rho_{1,0}},
    \end{eqnarray}
where we set its value equal to the value at the boundary
with the condensate at rest. It must have the same
value at the rarefaction wave boundary with the shelf
to the right of it:
    \begin{equation} \label{bound1}
    \overline{u}+2\sqrt{g_{11}\overline{\rho}_1}=2\sqrt{g_{11}\rho_{1,0}}.
    \end{equation}
The pressures of the components at the boundary
between them are equal, which gives the relation
    \begin{equation} \label{bound2}
    g_{11}\overline{\rho}_1^2=g_{22}\overline{\rho}_2^2.
    \end{equation}
Finally, after the passage through the dispersive shock
wave in the second component, the Riemann invariant $r_-^{(2)}$ retains its value, which gives the relation
    \begin{equation} \label{bound3}
    \overline{u}-2\sqrt{g_{22}\overline{\rho}_2}=-2\sqrt{g_{22}\rho_{2,0}}.
    \end{equation}
The three equations (\ref{bound1})-(\ref{bound3}) allow the densities of
the components on the shelves and the boundary
velocity to be found:
    \begin{equation} \label{tableau}
    \begin{split}
    \overline{\rho}_1=\frac1{\sqrt{g_{11}}}\left(\frac{\sqrt{g_{11}\rho_{1,0}}+\sqrt{g_{22}\rho_{2,0}}}
    {g_{11}^{1/4}+g_{22}^{1/4}}\right)^2,\qquad
    \overline{\rho}_2=\frac1{\sqrt{g_{22}}}\left(\frac{\sqrt{g_{11}\rho_{1,0}}+\sqrt{g_{22}\rho_{2,0}}}
    {g_{11}^{1/4}+g_{22}^{1/4}}\right)^2,
    \end{split}
    \end{equation}
    \begin{equation} \label{piston-vel}
    \overline{u}=\frac{2(g_{22}^{1/4}\sqrt{g_{11}\rho_{1,0}}-g_{11}^{1/4}\sqrt{g_{22}\rho_{2,0}})}{g_{11}^{1/4}+g_{22}^{1/4}}.
    \end{equation}
When inequality (\ref{pressure-cond}) holds, we have $\overline{u}>0$, i.e., the
boundary between the components moves to the right,
as it must be. The derived formulas are simplified considerably
if the nonlinear constants are equal, $g_{11}=g_{22}\equiv g$:
    \begin{equation} \label{g11=g22}
    \begin{split}
    \overline{\rho}_1=\overline{\rho}_2=\frac14(\sqrt{\rho_{1,0}}+\sqrt{\rho_{2,0}})^2,\qquad
    \overline{u}=\sqrt{g\rho_{1,0}}-\sqrt{g\rho_{2,0}}.
    \end{split}
    \end{equation}
In this case, the wave amplitude $\sqrt{\rho}$ on the plateau
between the rarefaction wave and the dispersive shock
wave is equal to the arithmetic mean of the component
amplitudes in the initial state, while the velocity of the
boundary between the components is equal to the difference
of the sound velocities in the initial states of
the components.

Having established the boundary values of the
parameters on both sides of the rarefaction wave and
the dispersive shock wave, we can turn to finding the
solutions for the waves themselves. We will begin with
the rarefaction wave. As we know, the Riemann invariant $r_+^{(1)}$ is constant along it (see (\ref{r1-pl})), so that Eqs. (\ref{rim2}) is satisfied identically for this invariant. The second
Riemann invariant $r_-^{(1)}$ depends only on the self-similar
variable $\zeta=x/t$, because our initial condition in the
form of a pressure jump at the boundary between the
components contains no parameter with the dimensions
of length. Therefore, Eq. (\ref{rim2}) is reduced to the
equation
    \begin{equation} \label{rim2-1}
    \frac{dr_-^{(1)}}{d\zeta}(v_-^{(1)}-\zeta)=0
    \end{equation}
with the obvious solution
    \begin{equation} \label{sol1}
    v_-^{(1)}=\frac14r_+^{(1)}+\frac34r_-^{(1)}=\frac{x}t.
    \end{equation}
Thus, along the rarefaction wave the Riemann invariants
are given by the expressions
    \begin{equation} \label{rim-inv1}
    r_+^{(1)}=2\sqrt{g_{11}\rho_{1,0}},\qquad r_-^{(1)}=\frac43\left(\frac{x}t-\frac12\sqrt{g_{11}\rho_{1,0}}\right).
    \end{equation}
Substituting them into (\ref{raref}), we will find the density of
the condensate and its flow velocity in the rarefaction
wave:
    \begin{equation} \label{raref2}
    \begin{split}
    \rho_1=\frac1{9g_{11}}\left(2\sqrt{g_{11}\rho_{1,0}}-\frac{x}t\right)^2,\qquad
    u_1=\frac23\left(\frac{x}t+\sqrt{g_{11}\rho_{1,0}}\right).
    \end{split}
    \end{equation}
The flow velocity becomes zero at the boundary with
the condensate at rest at $x_-^{(1)}=-\sqrt{g_{11}\rho_{1,0}}\cdot t$, i.e., the left
edge of the rarefaction wave propagates into the condensate at rest with a sound velocity $s_-^{(1)}=-\sqrt{g_{11}\rho_{1,0}}$ in
it. The right edge is joined with the shelf when $u_1=\overline{u}$, whence it follows that the velocity of the right edge is
    \begin{equation} \label{r-edge1}
    s_+^{(1)}=\frac{x_+^{(1)}}t=\frac{(2g_{22}^{1/4}-g_{11}^{1/4})\sqrt{g_{11}\rho_{1,0}}-3g_{11}^{1/4}\sqrt{g_{22}\rho_{2,0}}}
    {g_{11}^{1/4}+g_{22}^{1/4}}.
    \end{equation}
At $g_{11}=g_{22}=g$ velocities of the rarefaction wave
edges are
    \begin{equation} \label{edges-2}
    s_-^{(1)}=-\sqrt{g\rho_{1,0}}\qquad s_+^{(1)}=\frac{\sqrt{g}}{2}(3\sqrt{\rho_{2,0}}-\sqrt{\rho_{1,0}}).
    \end{equation}

If we attempt to find the wave between the plateau
and the second component at rest $\overline{\rho}_2$, $\overline{u}_2$ in the
dispersionless approximation, then we will arrive at
the multivalued solution indicated in Fig.~\ref{Step} by the
dashed line. This means that the dispersionless
approximation is inapplicable in this case, and allowance
for the dispersion leads to the replacement of the
multivalued solution by a region of fast oscillations
called a dispersionless shock wave. This region can be
roughly described as a modulated periodic solution of
the Gross-Pitaevskii equations with slowly changing
parameters. The equations describing the evolution of
the parameters can be derived by averaging the proper
number of conservation laws. This method of deriving
the modulation equations for nonlinear waves was
proposed by Whitham \cite{Whitham-65,Whitham-74}; for the Korteweg-de
Vries equation these equations were transformed by
him to a diagonal Riemann form analogous to the gasdynamic
equations (\ref{rim2}) The method developed by
Whitham was applied to the problem on the formation
of dispersive shock waves by Gurevich and Pitaevskii
in \cite{GurevichPitaevskii-73}. Since then the theory of dispersive shock
waves based on Whitham's method has been elaborated
greatly (see, e.g., the recent review \cite{ElHoefer-16}). In particular,
Whitham's modulation equations were derived
for periodic solutions of the Gross-Pitaevskii equation
(a nonlinear Schrodinger equation) in \cite{ForestLee-87,Pavlov-87}, the problem on the evolution of a step was analyzed in \cite{GurevichKrylov-87,ElGGK-95}, and the developed theory was applied to the
problems on the evolution of a condensate under the
action of a moving piston in \cite{HoeferAE-08,KamchatnovKorneev-10}. In our case, the
evolution of the second component to the right of the
interface between the components is reduced to the
problem on the flow of a condensate under the action
of a piston moving with velocity (\ref{piston-vel}). Therefore, the
results of the above papers are directly applicable to
our problem, and below we will provide the basic formulas
that describe the flow of the second condensate.

The periodic solution of the Gross-Pitaevskii
equations in a ``hydrodynamic form'' can be written as
    \begin{equation} \label{periodic}
    \begin{split}
    \rho&=\frac{1}{4g}(\lambda_4-\lambda_3-\lambda_2+\lambda_1)^2+\frac{1}{g}(\lambda_4-\lambda_3)(\lambda_2-\lambda_1)
    \mathrm{sn}^2(\sqrt{(\lambda_4-\lambda_2)(\lambda_3-\lambda_1)}\theta,m), \\
    u&=V-\frac{C}{g\rho},
    \end{split}
    \end{equation}
where
    \begin{equation}
    \begin{split}
    & \theta=x-Vt,\qquad V=\frac{1}{2}\sum_{i=1}^{4}\lambda_i,\qquad
    m=\frac{(\lambda_2-\lambda_1)(\lambda_4-\lambda_3)}{(\lambda_4-\lambda_2)(\lambda_3-\lambda_1)},\quad 0\leq m\leq1; \\
    & C=\frac{1}{8}(-\lambda_1-\lambda_2+\lambda_3+\lambda_4)(-\lambda_1+\lambda_2-\lambda_3+\lambda_4)
    (\lambda_1-\lambda_2-\lambda_3+\lambda_4);
    \end{split}
    \end{equation}
and the real parameters ?i $\lambda_i$ are ordered according to the
inequalities
    $$
    \lambda_1\leq\lambda_2\leq\lambda_3\leq\lambda_4.
    $$
As we see, the wave phase velocity $V$, amplitude $a=(\lambda_4-\lambda_3)(\lambda_2-\lambda_1)$, and background density $\rho_0=(\lambda_4-\lambda_3-\lambda_2+\lambda_1)^2/4$ through which the wave propagates
are expressed via these parameters. These parameters
are slow functions of $x$ and $t$ in a dispersive shock wave.
The periodic solution written in form (\ref{periodic}) has the
advantage that the parameters $\lambda_i$ are Riemann invariants,
and their evolution is defined by Whitham's
equations in a diagonal Riemann form (see~\cite{ForestLee-87,Pavlov-87})
    \begin{eqnarray} \label{Uisem}
    \frac{\partial\lambda_i}{\partial t}+v_i(\lambda_1,\lambda_2,\lambda_3,\lambda_4)\frac{\partial\lambda_i}{\partial x}=0,\quad i=1,2,3,4.
    \end{eqnarray}
Here, $v_i$ are the characteristic Whitham velocities:
    \begin{eqnarray}\label{whitham-eq}
    && v_1=\frac{1}{2}\sum_{i=1}^{4}\lambda_i-\frac{(\lambda_4-\lambda_1)(\lambda_2-\lambda_1)K}{(\lambda_4-\lambda_1)K-(\lambda_4-\lambda_2)E}, \\
    && v_2=\frac{1}{2}\sum_{i=1}^{4}\lambda_i+\frac{(\lambda_3-\lambda_2)(\lambda_2-\lambda_1)K}{(\lambda_3-\lambda_2)K-(\lambda_3-\lambda_1)E}, \\
    && v_3=\frac{1}{2}\sum_{i=1}^{4}\lambda_i-\frac{(\lambda_4-\lambda_3)(\lambda_3-\lambda_2)K}{(\lambda_3-\lambda_2)K-(\lambda_4-\lambda_2)E}, \\
    && v_4=\frac{1}{2}\sum_{i=1}^{4}\lambda_i+\frac{(\lambda_4-\lambda_3)(\lambda_4-\lambda_1)K}{(\lambda_4-\lambda_1)K-(\lambda_3-\lambda_1)E},
    \end{eqnarray}
where $K=K(m)$, $E=E(m)$ are elliptic integrals of
the first and second kinds, respectively. In the limit $m\rightarrow1$ ($\lambda_3\rightarrow\lambda_2$) a traveling wave transforms into a soliton solution against the background of a constant
density:
    \begin{equation}
    \begin{split}
    &\rho=\frac{1}{4}(\lambda_4-\lambda_1)^2-
    \frac{(\lambda_4-\lambda_2)(\lambda_2-\lambda_1)}
    {\mathrm{ch}^2(\sqrt{(\lambda_4-\lambda_2)(\lambda_2-\lambda_1)}\theta)},\\
    &\theta=x-\frac{1}{2}(\lambda_1+2\lambda_2+\lambda_4)t.
    \end{split}
    \end{equation}
In the other (small-amplitude) limit $m\rightarrow0$ ($\lambda_3\rightarrow\lambda_4$ or $\lambda_2\rightarrow\lambda_1$) the wave amplitude approaches zero, while
the density takes its background value. Significantly,
the pair of Whitham velocities in these limits transforms
(to within a constant coefficient) to the Riemann
velocities of the dispersionless limit. In particular,
in the soliton limit at $\lambda_3=\lambda_2$
    \begin{equation} \label{sol-limit}
    \begin{split}
    v_1(\lambda_1,\lambda_2,\lambda_2,\lambda_4)=\frac32\lambda_1+\frac12\lambda_4,\qquad
    v_4(\lambda_1,\lambda_2,\lambda_2,\lambda_4)=\frac32\lambda_4+\frac12\lambda_1,
    \end{split}
    \end{equation}
and in the small-amplitude limit needed for us at $\lambda_3=\lambda_4$
    \begin{equation} \label{small-limit}
    \begin{split}
    v_1(\lambda_1,\lambda_2,\lambda_4,\lambda_4)=\frac32\lambda_1+\frac12\lambda_2,\qquad
    v_2(\lambda_1,\lambda_2,\lambda_4,\lambda_4)=\frac32\lambda_2+\frac12\lambda_1.
    \end{split}
    \end{equation}
This means that the edges of the dispersive shock wave
are joined with the smooth solutions of the hydrodynamic
approximation, i.e., in our case with the shelf $\overline{\rho}_2$, $\overline{u}_2$ at the soliton edge and with the component at
rest $\rho_{2,0}$, $u_{2,0}=0$ at the small-amplitude edge.

    \begin{figure}[t] \centering
        \includegraphics[width=0.5\linewidth]{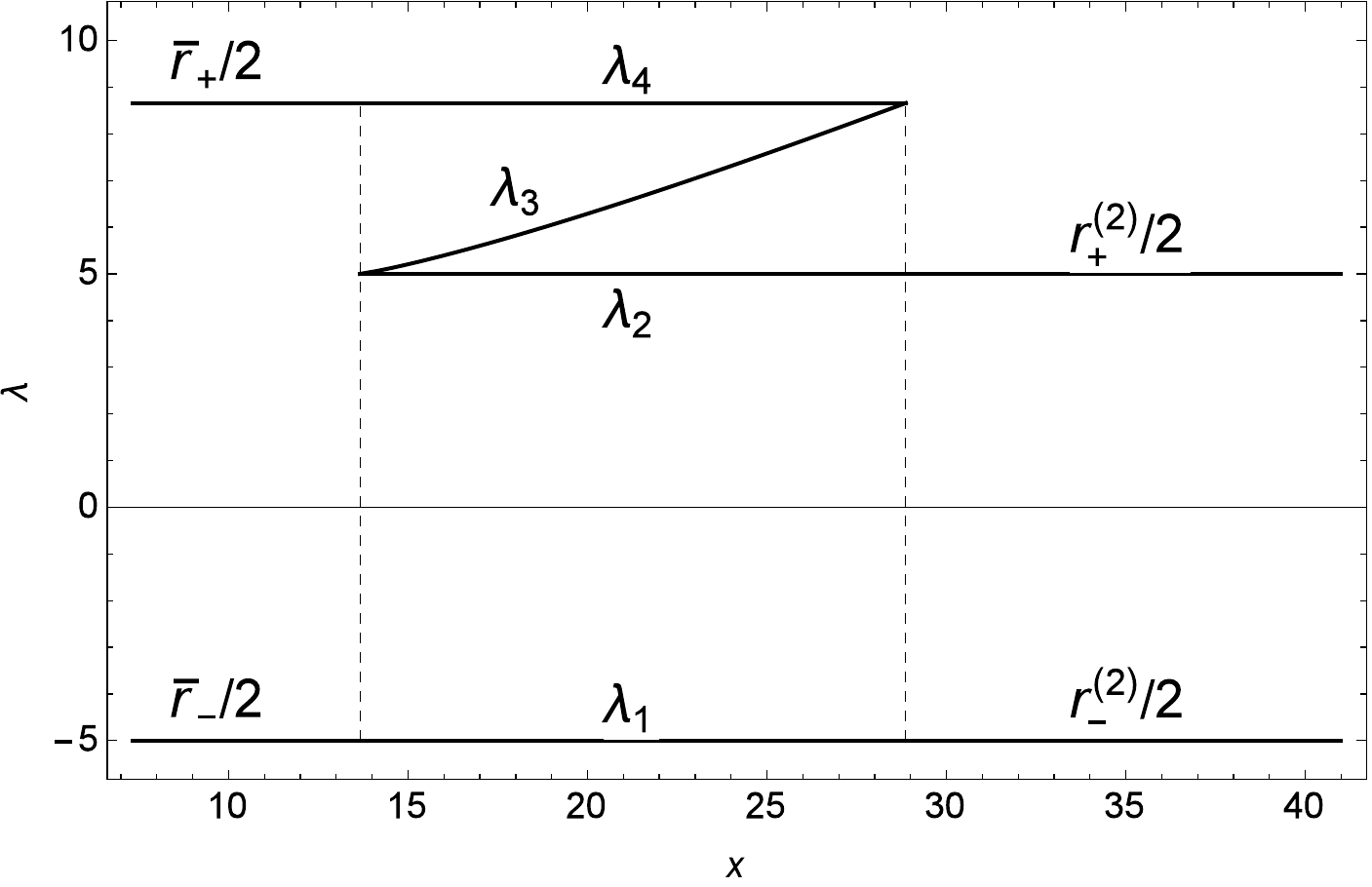}
        \caption{\textit{Graph for the Riemann invariants at $g_{11}=5$, $g_{22}=5$ and initial densities $\rho_{1,0}=15$, $\rho_{2,0}=5$ at $t=2$.}}
        \label{invariants}
    \end{figure}

\begin{figure}[t] \centering
        \includegraphics[width=0.6\linewidth]{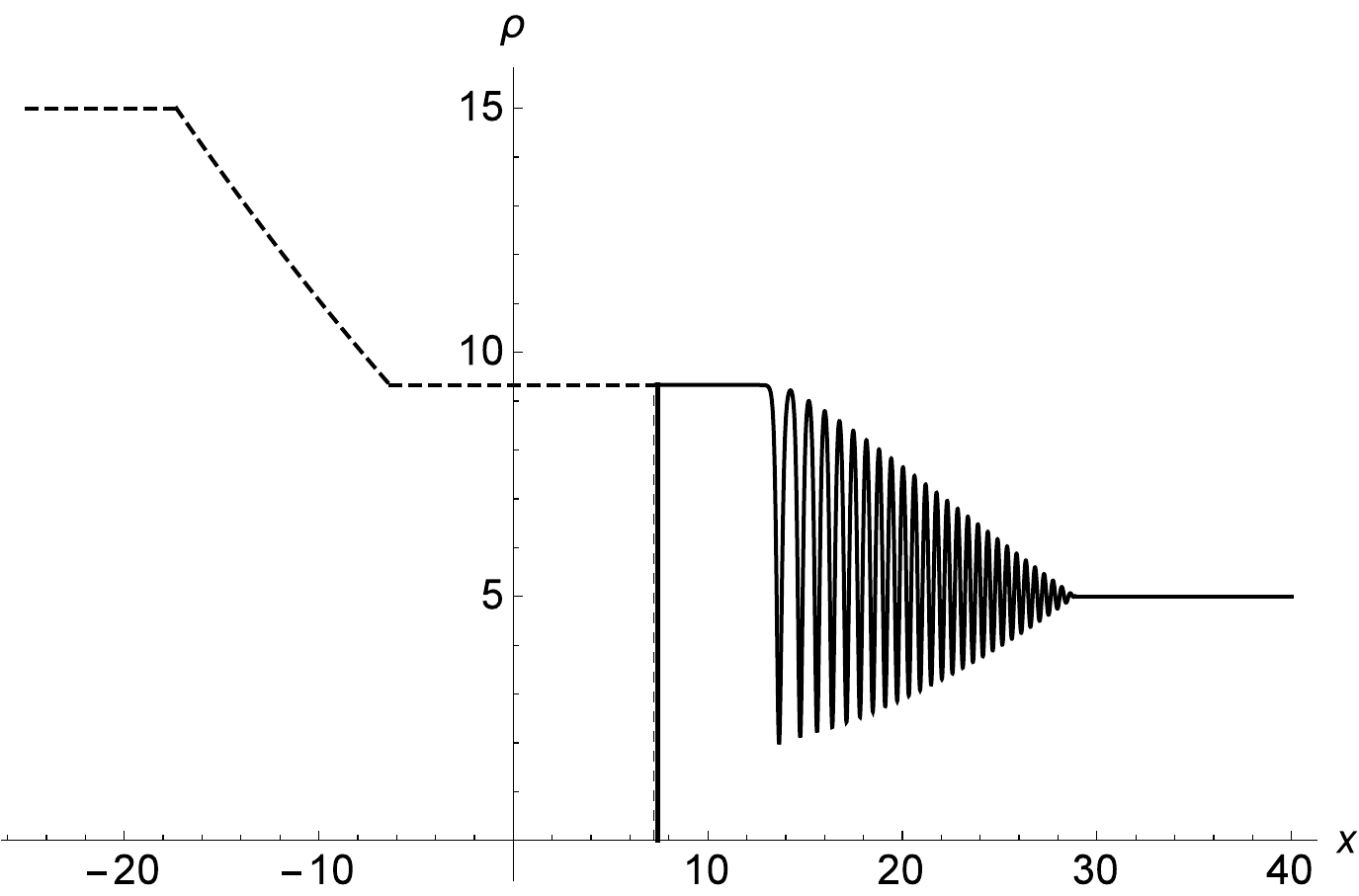}
        \caption{\textit{Analytical solution of the evolution of a step for immiscible components. The parameters are the same as those in Fig.~\ref{Step}. The dashed line indicates the left component.}}
        \label{StepAnalitic}
    \end{figure}

To elucidate the behavior of the solution of
Eqs. (\ref{Uisem}), we will again use the argument that the initial
condition contains no parameters with the dimensions
of length, so that the modulation parameters depend only on the self-similar variable $\zeta=x/t$. Therefore, Eqs. (\ref{Uisem}) are reduced to
    \begin{equation} \label{self-sim}
    \frac{d\lambda_i}{d\zeta}(v_i-\zeta)=0.
    \end{equation}
Hence it follows that only one Riemann invariant is
variable, while the three remaining ones must have
constant values. From the joining condition at the
edges of the dispersive shock wave we find that at the
soliton edge
    \begin{equation} \label{sol-limit2}
    \lambda_1=\overline{r}_-/2,\qquad \lambda_4=\overline{r}_+/2\qquad\text{at}\qquad \lambda_3=\lambda_2,
    \end{equation}
where $\overline{r}_{\pm}$ are the Riemann invariants of the dispersionless
theory that are defined by Eqs. (\ref{rim1}) and take the
values on the plateau bordering the soliton edge of the
dispersive shock wave. Similarly, at the small-amplitude
edge we find
    \begin{equation} \label{sol-limit2}
    \lambda_1=r_-^{(2)}/2,\qquad \lambda_2=r_+^{(2)}/2\qquad\text{at}\qquad \lambda_3=\lambda_4.
    \end{equation}
Thus, we arrive at the dependence of the Riemann
invariants on the spatial coordinate shown in Fig.~\ref{invariants} with the following constant Riemann invariants:
    \begin{equation} \label{rin-inv-const}
    \begin{split}
    \lambda_1&=-\sqrt{g_{22}\rho_{2,0}},\qquad \lambda_2=\sqrt{g_{22}\rho_{2,0}},\\
    \lambda_4&=\frac12\overline{u}+\sqrt{g_{22}\overline{\rho}_2}=\\
    &=\frac{2g_{22}^{1/4}\sqrt{g_{11}\rho_{1,0}}+
    (g_{22}^{1/4}-g_{11}^{1/4})\sqrt{g_{22}\rho_{2,0}}}{g_{11}^{1/4}+g_{22}^{1/4}}.
    \end{split}
    \end{equation}
In contrast, the dependence of the variable Riemann
invariant $\lambda_3$ on $\zeta=x/t$ is implicitly defined by the
equation
    \begin{equation} \label{la3}
    v_3\left(-\sqrt{g_{22}\rho_{2,0}},\sqrt{g_{22}\rho_{2,0}},\lambda_3,\tfrac12\overline{u}+\sqrt{g_{22}\overline{\rho}_2}\right)=\frac{x}t.
    \end{equation}
Substituting the Riemann invariants found into the
periodic solution (\ref{periodic}) gives the coordinate dependence
of the condensate density at a fixed time. It is
illustrated in Fig.~\ref{StepAnalitic} for the same values of the parameters
at which our numerical calculation shown in Fig.~\ref{Step}. was made. As we see, there is good agreement of
the analytical results with the numerical ones.

The derived formulas also give analytical expressions
for the velocities of the edges of the dispersive
shock wave. The soliton edge moves with the velocity
    \begin{equation} \label{s-}
    \begin{split}
    s_-^{(2)}& = v_3\left(\lambda_1,\lambda_2,\lambda_2,\lambda_4\right)=\frac12(\lambda_1+2\lambda_2+\lambda_4)=\\
    &=\frac{g_{22}^{1/4}(\sqrt{g_{11}\rho_{1,0}}+\sqrt{g_{22}\rho_{2,0}})}{g_{11}^{1/4}+g_{22}^{1/4}},
    \end{split}
    \end{equation}
while the velocity of the small-amplitude edge is
    \begin{equation} \label{s+}
    \begin{split}
    s_+^{(2)}&= v_3\left(\lambda_1,\lambda_2,\lambda_4,\lambda_4\right)=\\
    &=\lambda_4+\frac{\lambda_1+\lambda_2}2+\frac{2(\lambda_4-\lambda_2)(\lambda_4-\lambda_1)}{2\lambda_4-\lambda_1-\lambda_2}=\\
    &=\frac{2(\overline{u}+\sqrt{g_{22}\rho_{2,0}})^2-g_{22}\rho_{2,0}}{\overline{u}+\sqrt{g_{22}\rho_{2,0}}}.
    \end{split}
    \end{equation}
When the nonlinear constants are equal, $g_{11}=g_{22}=g$, these velocities are
    \begin{equation} \label{speeds}
    \begin{split}
    s_-^{(2)}=\frac{\sqrt{g}}2(\sqrt{\rho_{1,0}}+\sqrt{\rho_{2,0}}),\qquad
    s_+^{(2)}=\frac{\sqrt{g}(2\rho_{1,0}-\rho_{2,0})}{\sqrt{\rho_{1,0}}}.
    \end{split}
    \end{equation}
These values also agree well with the results of our
numerical calculation.

\section{Conclusions}

The results of this paper show that the expansion
dynamics of a two-component condensate is distinguished
by a great diversity compared to the one-component
case. First, the initial states of the two-component
condensate can form various configurations in a
nonuniform trap field, and we constructed a phase
diagram of these states by refining the previous studies
of other authors. Second, the simple self-similar
ansatz that was successfully used in the theory of the
expansion of a one-component condensate now has a
limited applicability, describing in general terms the
expansion of the two-component condensate only far
from the transition line from the miscible components
to the immiscible ones. Finally, third, if the components
are immiscible, then the expansion regimes with the generation of dispersive shock waves are possible.
This can be of interest both for the analysis of condensate
flows in specific experimental conditions and for
nonlinear physics as a whole. The results obtained in
this paper allow one to predict the main characteristics
of the expansion dynamics and to estimate the main
parameters of the emerging wave structures.

\section*{Acknowledgments}

This work was supported by the Russian Foundation
for Basic Research (project no. 16-01-00398).

\end{document}